\documentclass{aa}  

\usepackage{graphicx}
\usepackage{txfonts}
\usepackage{xcolor}

\usepackage[colorlinks=true, citecolor=blue]{hyperref}
\begin{document}

   \title{On the nature of VX~Sagitarii}

   \subtitle{Is it a T\.ZO, a RSG or a high-mass AGB star?}

   \author{H. M. Tabernero \inst{1,2} \thanks{Based (partly) on data obtained with the STELLA robotic telescopes in Tenerife, an AIP facility jointly operated by AIP and IAC.}$^,$\thanks{Based (partly) on observations collected at the European Southern Observatory under ESO programme 298.D-5004(A).}
   \and
R. Dorda \inst{2,3,4}
\and
I. Negueruela \inst{5}
\and
E. Marfil \inst{6}  
          }

   \institute{Instituto de Astrof{\'i}sica e Ci{\^e}ncias do Espa\c{c}o, Universidade do Porto, CAUP, Rua das Estrelas, 4150-762 Porto, Portugal
              \email{htabernero@astro.up.pt}
         \and
      Departamento de F\'{\i}sica, Ingenier\'{\i}a de Sistemas y Teor\'{\i}a de la Se\~nal, Universidad de Alicante, Carretera de San Vicente s/n, E03690, San Vicente del Raspeig, Spain
\and
Instituto de Astrof\'{i}sica de Canarias (IAC), 38205 La Laguna, Tenerife, Spain
\and
Universidad de La Laguna (ULL), Departamento de Astrof\'{i}sica, 38206 La Laguna, Tenerife, Spain
\and
Departamento de F\'{\i}sica Aplicada, Facultad de Ciencias, Universidad de Alicante, Carretera de San Vicente s/n, E03690, San Vicente del Raspeig, Spain
\and
Departamento de F{\'i}sica de la Tierra y Astrof{\'i}sica \& IPARCOS-UCM (Instituto de F\'{i}sica de Part\'{i}culas y del Cosmos de la UCM), Facultad de Ciencias F{\'i}sicas, Universidad Complutense de Madrid, 28040 Madrid, Spain
             }
   \date{Received August 22, 2020; accepted MM DD, YYYY}

  \abstract
  {}
  {We present a spectroscopic analysis of the extremely luminous red star VX~Sgr based on high-resolution observations combined with AAVSO light curve data. Given the puzzling characteristics of VX~Sgr, we explore three scenarios for its nature: a massive red supergiant (RSG) or hypergiant (RHG), a Thorne \.{Z}ytkow object (T\.ZO), and an extreme Asymptotic Giant Branch (AGB) star.  }
  {Sampling more than one whole cycle of photometric variability, we derive stellar atmospheric parameters by  using state-of-the-art PHOENIX atmospheric models. We compare them to optical and near infrared spectral types. We report on some key features due to neutral elemental atomic species such as \ion{Li}{i}, \ion{Ca}{i}, and \ion{Rb}{i}. }
   {We provide new insights into its luminosity, evolutionary stage as well as its pulsation period.  Based on all the data, there are two strong reasons to believe that VX~Sgr is some sort of extreme AGB star. Firstly, its Mira-like behaviour during active phases. VX~Sgr shows light variations with amplitudes much larger than any known RSG and clearly larger than all RHGs. In addition, it displays Balmer line emission and, as shown here for the first time, line doubling of its metallic spectrum at maximum light, both characteristics typical of Miras. Secondly, unlike any known RSG or RHG, VX~Sgr displays strong \ion{Rb}{i} lines. In addition to the photospheric lines that are sometimes seen, it always shows circumstellar components whose expansion velocity is compatible with that of the OH masers in the envelope, demonstrating a continuous enrichment of the outer atmosphere with s-process elements, a behaviour that can only be explained by third dredge up during the thermal pulse phase.}
   {}

   \maketitle
%

\section{Introduction}
\label{intro}

VX~Sgr is an extremely luminous red star that exhibits fluctuating spectral veiling and semi-regular photometric and spectroscopic variations \citep{all1951} shaped by quiescent and active variation phases. Due to the presence of SiO, OH, and H$_{2}$O masers, it is believed to be wrapped inside a large molecular envelope. However, its nature is still unclear, as its initial mass and evolutionary state are poorly constrained. For instance, while most of the analyses available in the literature consider VX~Sgr as a high-mass ($M\gtrsim10\:$M$_{\odot}$) red supergiant (RSG) or even as a red hypergiant \citep[RHG; see][]{sch2006}, other authors \citep[e.g.][]{chi2010} point instead that the atmosphere of VX~Sgr appears to be more similar to an oxygen-rich asymptotic giant branch star \citep[AGB; see][]{her2005}, with $M\sim4$\,--\,$10\:$M$_{\odot}$.\\

The high-mass of VX~Sgr is supported by its high luminosity, which critically depends on the distance adopted. According to the distance values determined during the last decades \citep[see the briefing presented by][and their Table~1]{che2007}, its luminosity has been estimated in the range $-9.3\:{\rm mag} \leq M_{\mathrm{ bol}} \leq -7.8\:$mag \citep{loc1982,che2007,arr2013,chi2010,liu2017}.  Therefore, according to most luminosity estimates, VX~Sgr seems to be brighter than the most luminous AGB stars observed to date \citep[$M_{\mathrm{bol}}\sim-8\:$mag;][]{gro2009,gar2009}. However, it remains uncertain whether RSGs can actually reach such high luminosities. For example, Geneva evolutionary tracks \citep{eks2012} predict that RSGs cannot simultaneously attain the high luminosity and the low effective temperature ($T_{\rm eff}$) seen in VX~Sgr, although such stars are known to exist (i.e., RHGs). Moreover, RSGs with high initial masses ($25<M/$M$_{\odot}<40$) present luminosities up to $M_{\mathrm{bol}}\sim-9.0\:\:$mag \citep{dav2018}, but they quickly evolve back towards higher $T_{\rm eff}$.\\

After a careful analysis of the outer atmosphere of VX~Sgr through spectro-interferometric imaging, \citet{chi2010} concluded that it qualitatively resembles a Mira-like star. In addition, they found that its H$_{2}$O layers resemble those of an AGB, although they stressed that their Mira model takes stellar parameters (i.e., $M=1.2\:$M$_ {\odot}$, $M_{\mathrm{ bol}}\sim-4.5\:$mag) that are not consistent with the expected parameters for VX~Sgr. Similarly, the spectral and photometric variations of VX~Sgr present more analogies to AGB stars with Mira-like variability than to most RSGs. Its visual magnitude exhibits peak-to-peak changes of up to $6\:$mag (see AAVSO\footnote{The American Association of Variable Star Observers; see their light curve generator at \url{https://www.aavso.org/}}), whereas typical RSGs show changes no larger than $2\:$mag \citep{kis2006}. Even extreme RHGs \citep[e.g. S~Per, UY~Sct, and VY~CMa;][]{sch2006} show a maximum peak-to-peak variation of only $4\:$mag, according to AAVSO photometric data. In the $I$-band, RSGs present variation amplitudes smaller than $0.45\:$mag, while AGB stars have larger amplitudes \citep{gro2009}. In contrast, VX~Sgr has a peak-to-peak variation of $1.2\:$mag \citep{loc1982} in that  photometric band. We also note that the spectral variation of VX~Sgr is unique among RSGs since it spans around 6 spectral subtypes \citep[from M4 to M10;][]{hum1974a,loc1982}. Most Galactic RSG have variations of one subtype or less \citep{whi1978}, whereas the RHGs change by no more than 2-3 subtypes.\\

The veiling of VX~Sgr \citep{hum1974a}, which manifests itself as a weakening of all spectral features, probably as a result of the scattering induced by the circumstellar shell \citep{mas2009}, provides further evidence that VX~Sgr does not match the hallmarks of RSGs. Veiling has been reported for Mira stars \citep[e.g.][]{mer1962} and extreme RHGs such as S~Per or VY~CMa, but not for RSGs as far as we know. In fact, the sample of approximately 600 stars studied by \citet{dor2016b} included only two veiled stars,  S~Per and UY~Sct, which were already well-known RHGs.\\

After revisiting the information available in the literature about VX~Sgr, we found a significant spread in distance determinations, with some authors placing VX~Sgr much closer to us than usually assumed, which in turn translates into lower luminosity estimations, compatible with those of an AGB star. However, many features of VX~Sgr do not match those of an AGB star. Firstly, when not veiled, the spectrum of VX~Sgr has been classified as very luminous \citep[luminosity class Ia;][]{hum1974a} independently from the distance adopted. Secondly, the circumstellar envelope of VX~Sgr is not expected for a typical AGB~star. For example, \cite{ric2012} estimated the radius of the H$_{2}$O maser cloud in VX~Sgr to be an order of magnitude larger than those around known AGB stars. \cite{arr2013}, based on data from \cite{chi2010} and \cite{su2012}, also found that the SiO masers of VX~Sgr are located at \textasciitilde3 stellar radii, while the SiO masers in AGB stars are mostly located at \textasciitilde2 stellar radii. On the other hand, these results for VX~Sgr are consistent with those obtained for other RHGs, e.g. S~Per \citep{ric2012} and AH~Sco \citep{arr2013}. Finally, we note that VX~Sgr is by no means unique, in the sense that other RHGs share some of the features of VX~Sgr. In addition, their extremely high luminosities, derived from accurate distances, are close to that of VX~Sgr \citep{sch2006,arr2013}.\\

In principle, some of the characteristics of VX~Sgr, such as its high luminosity and its variability, could be more easily explained in a binary scenario. However, \citet{kam2005} did not find a long secondary period in the AAVSO light curve caused by a tentative stellar companion to VX~Sgr.  Furthermore, they found that the circumstellar envelope of VX~Sgr is nearly spherical. This result was later confirmed by \citet{yoo2018}, thus discarding signs of interaction with potential companions.\\

At this point, a question arises: what is then the nature of VX~Sgr? To answer it, we adopted a strategy different from previous studies. For several decades, studies of VX~Sgr have focused on either its circumstellar envelope or its structure, since these features can be resolved by interferometric methods. In fact, VX~Sgr has not been specifically studied through optical spectroscopy since the 80's. Spectroscopic analysis tools and spectral classification techniques have significantly improved since then, and now it is possible to explore new scenarios for this fundamentally challenging star. In this work, we discussed the possibilities of VX~Sgr being a RSG/RHG, an AGB, or  even a Thorne–\.Zytkow object (T\.ZO). These purely hypothetical stars are red giants or supergiants that have engulfed their companion neutron star inside their cores. Due to this circumstance, they can theoretically reach the luminosities of RSGs. In addition, they will produce an overabundance of some chemical species such as  Li \citep{pod1995}, as well as Ca, or Rb \citep[see ][]{bie1994}. Although their existence was first suggested by \cite{tho1977}, very few candidates have been proposed ever since. One candidate is HV~2112, an extremely luminous OH/IR star located in the Small Magellanic Cloud \citep{lev2014}. However, \citet{bea2018} only found signs of an overabundance of Li, and concluded that HV~2112 is a thermally pulsing AGB star rather than a T\.ZO.\\

This manuscript is divided into seven different sections. In Sect.~\ref{obs} we describe how the observations were obtained. In Sect.~\ref{physparam} we describe how the star was characterised. We report on different spectral features of the spectrum of VX~Sgr in Sect.~\ref{sp_feat}. We analyse the historical light curve of VX~Sgr in Sect.~\ref{period}. We discuss our findings in Sect.~\ref{discussion}. Finally, the conclusions of this manuscript can be found in Sect~\ref{conclusions}.

\section{Observations}
\label{obs}

We observed VX~Sgr on 11 different epochs  between April 2016 and June 2018. We used the STELLA echelle spectrograph (SES) at the 1.2~m robotic telescope STELLA \citep[Iza\~na Observatory, Tenerife, see][]{web12} and the Ultraviolet and Visual Echelle Spectrograph (UVES) at the Very Large Telescope (VLT) \citep[Paranal Observatory, see][]{dek00}. Most of our programme was done using SES (8 epochs), while only 3 epochs were obtained with UVES, under ESO proposal 298.D-5004(A). The spectral coverage of SES ranges from $3\,900$ to $8\,700\:$\AA{} with a spectral resolution of $55\,000$, whereas the red arm of UVES covers from $6\,800$ to $11\,000\:$\AA{} with a spectral resolution of about $110\,000$. The UVES data were reduced using the corresponding ESO reduction pipeline\footnote{https://www.eso.org/sci/software/pipelines/uves/uves-pipe-recipes.html}, whereas the SES data are automatically reduced on-site. We give further details of these spectroscopic observations in Table~\ref{epochs}. Finally, as  complementary data to our spectroscopic observations, we downloaded the photometric measurements registered by  AAVSO. These data covers a time-span of about 85 years from 1935-05-25 (BJD~$=$~2427948) to 2020-07-27 (BJD~$=$~2459055), which corresponds to the whole historical record available for VX~Sgr in the AAVSO database.

\begin{table*}
\tiny
\centering
\begin{tabular}{cccccccccc}
\hline

Date       & BJD & Instrument  &    t$_{exp}$  &  RV           & $T_{\mathrm{eff}}$ & $\zeta$ &  \multicolumn{2}{c}{SpT}   & Remark\\
           &    &              &  [s]          & [km~s$^{-1}$] & [K]               & [km~s$^{-1}$] & Optical               & Near infrared         &   \\
\hline
\hline
\noalign{\smallskip}
2016-04-04  & 2457482 & SES & 3600  &   $-4.3\pm0.2$          & 3408~$\pm\:$37  &  4.19~$\pm\:$0.36 &   M7  &   M8  & Rb double lined\\  
2016-05-21 & 2457529 & SES  & 3600 &  $-4.5\pm0.2$          & 3360~$\pm\:$32  &  4.72~$\pm\:$0.54  &  M8  & M8.5  & Rb double lined\\
2016-09-22 &  2457653 & SES & 6000  &   $-5.0\pm0.4$         & 3523~$\pm\:$25  &  6.87~$\pm\:$0.32 &   M5.5  &   M7   & H in emission\\ 
2016-10-25 & 2457686 & UVES &  10.5   &  $-4.9\pm0.2$          & 3471~$\pm\:$13  &  9.19~$\pm\:$0.86 &   M4.5  &   M7 & -- \\
2017-02-12 & 2457796 & UVES  &  2.5  &  $-5.2\pm0.1$           & 3318~$\pm\:$10  & 10.18~$\pm\:$0.80 & M5.5  &   M7.5   & -- \\
2017-03-04 & 2457816 & UVES  &  8.5  & $-5.2\pm0.1$           & 3261~$\pm\:$68  & 10.36~$\pm\:$0.90 & M5.5  &   M7.5  & -- \\
2017-04-02 & 2457845 & SES   & 3000 &  $-4.7\pm0.2$           & 3347~$\pm\:$24  & 6.54~$\pm\:$0.55 &  M6    &   M8  & -- \\
2017-05-17 & 2457890 & SES   & 3000 & $-4.3\pm0.2$           & 3273~$\pm\:$29  & 6.26~$\pm\:$0.50 &  M6    &   M8   & -- \\
2017-06-29 &2457933 & SES  & 3000  & $-4.1\pm0.2$           & 3294~$\pm\:$37  & 6.49~$\pm\:$0.56 &  M6.5    &   M8   & -- \\
2017-08-06 &  2457971& SES & 6000 & $-3.6\pm0.2$           & 3272~$\pm\:$36  & 5.36~$\pm\:$0.75 &   M7 &   M8.5 & -- \\
2018-06-06 & 2458275 & SES & 3000 & $-3.3\pm0.3$          & 3562~$\pm\:$19  & 6.24~$\pm\:$0.38 &   M3 &   M4  & H in emission; Doubled atomic lines.\\
\hline
\end{tabular}
\caption{Epochs of observation, instrument used to acquire each observation, total exposure times for each epoch (t$_{exp}$), radial velocities (RV), Specral types (SpT), and physical parameters for VX~Sgr ($T_{\rm eff}$, $\zeta$). $\log{g}$ is fixed to $-0.5\:$dex whereas [M/H] is also fixed to $0.0\:$dex (see text for details). The accuracy of the spectral classification is one subtype.}
\label{epochs}
\end{table*}

\section{Physical parameters}
\label{physparam}
\subsection{Distance to VX~Sgr}
\label{dist}

The first distance estimation was inferred by \cite{hum1972},  who assumed that VX~Sgr belongs to the OB association Sgr~OB1, located at 1.5-1.7~kpc \citep{mor1953, hum1975}.  Later, \cite{hum1974a} derived a distance of 0.8~kpc using the photometry given by \cite{lee1970}. A few years later \cite{loc1982} reported a photometric distance of 1.5~kpc. The trigonometric parallaxes obtained by \textit{Hipparcos} led to distances even shorter than any previous estimations: $\sim0.3\:$~$\pm$~$0.2$~kpc \citep{newhip}. We believe that these measurements are, to a large extent, highly unreliable and not statistically significant \citep[e.g.][]{pou2003}. This is caused by the combination of a large angular diameter \citep[$\sim8.8$~mas, see ][]{mon2004,chi2010}, far larger than the parallax calculated, and the presence of convective bright spots which vary over time \citep{chi2011,pas2011}. More recently, the Gaia Data Release 2 \citep[GDR2;][]{gaiaDR2} has provided a parallax of $\pi=0.79\pm0.23$ for VX~Sgr, and, from this, \citet{bai2018} calculated a distance of $1.36_{-0.41}^{+1.02}\:$kpc, a value with a very high uncertainty, specially toward its upper limit.\\

In the past twenty years, with the exception of the GDR2, all the distances to VX~Sgr have been calculated through the masers in its circumstellar structure. \cite{mar1998}, \cite{mur2003} and \cite{xu2018} used H$_2$O masers, and they all obtained similar distances: $1.7\pm0.3$, $1.8\pm0.5$ and $1.56^{+0.11}_{-0.10}\:$kpc respectively.  Likewise, \cite{che2007} and \cite{su2018} derived a distance (using SiO masers) of $1.57\pm0.27$ and $1.10\pm0.11\:$kpc, respectively. All these values, except that by \cite{su2018}, are in excellent agreement. In fact, \cite{xu2018} showed that all the astrometric parameters (parallax, radial velocity, and proper motions) for VX~Sgr are in good agreement with those found in the literature for Sgr~OB1.

\subsection{Radial Velocities}
\label{Velocities}

We calculated the RVs following the cross-match algorithm described in \citet{pepe2002}, which computes the cross-correlation function (CCF) by using line masks around previously selected spectral features. Since the spectrum of VX~Sgr is dominated by TiO molecular bands that cover both the optical and the near infrared, we decided to use them to compute the RVs by employing a list of TiO vibro-rotational transitions from the Vienna Atomic Line Database\footnote{\url{http://vald.astro.uu.se/}} \citep[VALD3;][]{vald3}. Among those lines, we selected only those transitions with wavelengths accurately measured in the laboratory. In particular, we sampled the CCF from $-$100 to 100~km~s$^{-1}$ with a step of 0.5~km~s$^{-1}$, using masks that are 1~km~s$^{-1}$ wide and proportionally weighted to their normalised intensity with respect to the stellar continuum. We assigned each individual weight according to the VALD3 {\tt extract stellar} option using the coolest model available in the database ($T_{\rm eff}$~$=$~3500~K and $\log{g}$~$=$~0.0~dex). Finally, we fitted a Gaussian profile to each individual CCF to obtain the corresponding RVs. Uncertainties in the RVs were calculated by means of the scheme described in \citet{zuck03} using the implementation of \citet{bla14}.\\

\begin{figure}
   \centering
      \includegraphics[width=\columnwidth]{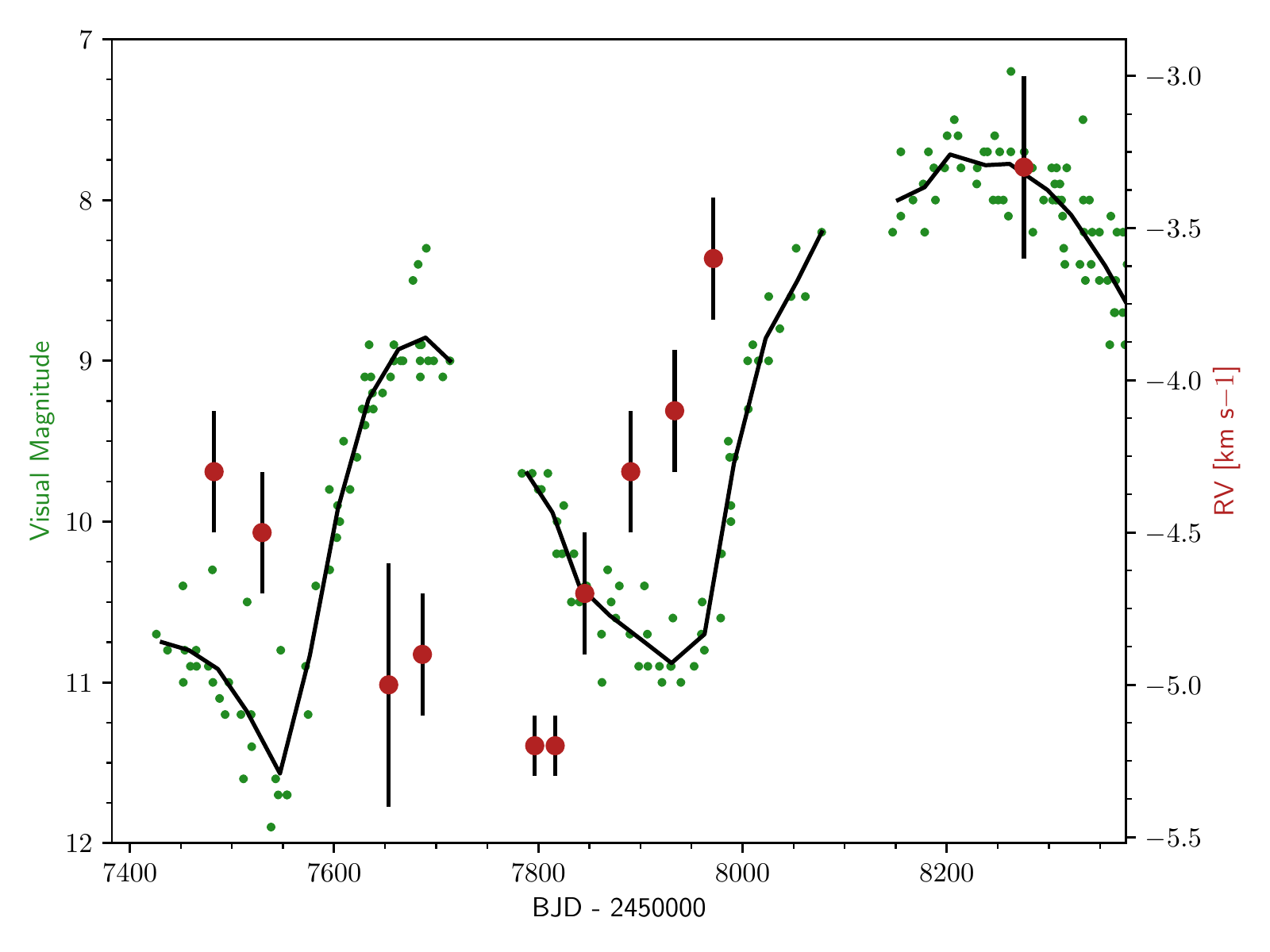}
   \caption{Light curve of VX~Sgr between February 2016 and October 2018, with our spectroscopic data superimposed. Green dots indicate individual photometric points, measured in visual magnitudes (left-$y$ axis). The black line represents the smoothed light curve (i.e. 30~d binning).  Red circles indicate the radial velocity of the corresponding spectrum (right-$y$ axis).}
   \label{recent_rv}
\end{figure}

The uncertainties of our calculated radial velocities are smaller than the standard deviation of the individual measurements. Thus, we compared the radial velocities against the AAVSO light curve in Fig.~\ref{recent_rv}. We found that our values are shifted with respect to the light curve. This behaviour has already been reported for pulsating Mira stars by \citet{jor2016} and supported by the theoretical predictions of \citet{lil2018}. According to these two studies, the maximum light occurs when the atmosphere reaches its maximum contraction, shortly after the maximum velocity is reached. This is precisely the behaviour we observed in Fig.~\ref{recent_rv}. Therefore, the velocity range that we have observed in the individual measurements (see Table~\ref{epochs}) is only due to the Mira-like behaviour of VX~Sgr.

\subsection{Stellar atmospheric parameters}
\label{StePar}

 We used the automatic tool {\sc SteParSyn} \citep{tab18} to calculate the stellar parameters of VX~Sgr. The latest version of {\sc SteParSyn} relies on {\tt emcee} \citep{emcee}, a Markov Chain Monte Carlo (MCMC) method used to fully sample the underlying distribution of the stellar atmospheric parameters. {\sc SteParSyn}  has been used to infer stellar atmospheric parameters using atomic features \citep[see e.g.][]{lohr18,neg18,tab18,alo19}.  Unfortunately, the optical spectrum of VX~Sgr is dominated by molecular absorptions. In consequence, we must rely on an alternative indicator to the atomic lines; in particular, the TiO bands are a good proxy to model a star like VX~Sgr \citep[see e.g.][]{gar2006,gar2009}. Thus, we selected the TiO band system at $7\,050\:$\AA{}, as these molecular bands dominate the optical spectrum of VX~Sgr. In addition, this TiO-band system is available in both the UVES and SES wavelength ranges.\\

We employed the state-of-the-art PHOENIX-ACES model grid \citep{phoenixaces} to fit our observations. Each synthetic spectrum in the PHOENIX grid is characterised by its effective temperature ($T_{\mathrm{eff}}$), surface gravity ($\log{g}$), and metallicity ([M/H]). Among the spectra available in the grid, we selected those with $T_{\mathrm{eff}}$ between $2\,300$ and $4\,000\:$K. Metallicity and $\log{g}$ are known to be degenerate, but in almost all cases our best fit resulted in metallicity very close to solar and $\log{g}$ below zero. Therefore we fixed $\log{g}$ to $-0.5\:$dex, following the prescription described in \citet{zam2014,per2017}, and fixed [M/H] to solar, which is also the expected metallicity for young stars in the disk of the Milky Way. In addition, we took into account the resolution and any remaining sources of broadening by means of a Gaussian kernel spanning from 1~to 20\:km~s$^{-1}$ in FWHM. The Gaussian broadening accounts for both the macro-turbulence and the instrumental broadening.  Given its size, we do not expect VX~Sgr to have a large rotational velocity, and therefore any broadening that is not explained by the instrumental profile can be explained to a large extent by macroturbulence ($\zeta$). Finally, our derived stellar parameters and their uncertainties can be found in Table~\ref{epochs} whereas the best fits to the data are shown in Fig.~\ref{best_fit}.

\subsection{Luminosity}
\label{lumos}

Several authors have inferred the luminosity of VX~Sgr (as $M_{\mathrm{bol}}$) by means of different methods  alongside the distance measurements available to them. \cite{loc1982} derived $M_{\mathrm{bol}}$ from their photometric data. Assuming a distance of $1.5\:$kpc, they estimated a $M_{\mathrm{bol}}$  in the range $-8.4\:{\rm mag} \leq M_{\mathrm{ bol}} \leq -7.8\:$mag. \cite{chi2010} used interferometric observations to calculate an angular diameter of $\Theta\:=\:8.82\:\pm\:0.5\:$mas.  They  assumed an $T_{\mathrm{eff}}$ in the range from $3\,200$ to $3\,400\:$K and a distance of $1.7\:$kpc to obtain a $M_{\mathrm{bol}}=-8.4\:$mag. \cite{arr2013} later calculated the bolometric flux of VX~Sgr through their spectro-interferometric observations. They used this value along with the angular diameter from \cite{chi2010} and a distance of $1.57\pm0.27\:$kpc from \citet{che2007}. Their results were $M_{\mathrm{bol}}=-9.3\:$mag and $T_{\mathrm{eff}}\sim3\,750\:$K. Finally, \cite{liu2017} modelled its SED from the optical to the infrared and obtained $M_{\mathrm{bol}}=-9.1\:$mag and $T_{\mathrm{eff}}\sim3\,150\:$K.\\

The data currently available give room to two different methodologies to calculate the bolometric magnitude. On the one hand, a photometric method can be used, by calculating the bolometric magnitude through 2MASS magnitudes, but this compels us to assume an extinction ($A_{V}$), which is by no means an easy task. Firstly, the calculation of $A_{V}$ to VX~Sgr requires prior knowledge of the intrinsic colours of this star. In addition, the spectral variability of VX~Sgr is extreme, ranging from M3 up to M10. Very luminous (Ia)~M stars are also particularly scarce. In consequence, the calibrations available \citep[e.g.][]{eli1985} extend only down to spectral type M4.  Alternatively, we calculated the luminosity  of VX~Sgr by means  of the following equation:\\

\begin{equation}
M_{\mathrm{bol}}=-10\log\left(\frac{T_{\mathrm{eff,VX}}}{T_{\mathrm{eff},\odot}}\right)-5\log\left(\frac{\Theta d}{2 R_{\odot}}\right)+M_{\mathrm{bol}}^{\odot}\, .
\label{ec1}
\end{equation}

The expression for $M_{\rm bol}$ given by Equation~\ref{ec1} depends on three observable quantities: angular diameter ($\Theta$), distance ($d$), and effective temperature ($T_{\rm eff}$). The measured angular sizes of VX~Sgr can be found in the literature: $8.7$~$\pm$~$0.3$ and $8.82$~$\pm$~$0.5$~mas, from \citet{mon2004} and \citet{chi2010}, respectively. We averaged these two values into a single angular diameter of  $8.76$~$\pm$~$0.4$~mas. In addition, we computed an average distance from the literature values listed in Section~\ref{dist} by weighting each distance value with the inverse of its uncertainty, which yielded $1.44\pm0.19\:$kpc. Finally, our measurements showed an average effective temperature of $3\,370\:\pm\:100$K. We used these three quantities (namely $\Theta$, $d$, and $T_{\rm eff}$) to calculate a $M_{\mathrm{bol}}$ of $-8.6$~$\pm$~$0.6$~mag. The luminosity thus calculated is above the estimations for the most luminous AGB stars   ($M_{\mathrm{bol}}\-8.2\:$mag) calculated by \citet{doh2015}. In addition, our calculated $M_{\rm bol}$ is brighter than the maximum luminosity observed in a confirmed AGB \citep[$M_{\mathrm{bol}}\sim-8\:$mag, see][]{gro2009, gar2009}. 
However, its high luminosity in itself does not rule out an AGB nature for VX~Sgr. In fact, theoretical models for stars at the end of the AGB and super-AGB phases have convergence issues \citep[see, e.g.,][]{lau2012} that presently make impossible to produce any theoretical predictions to compare with the observational data. 

\subsection{Association membership}
\label{asoc_mem}

Traditionally, VX~Sgr has been considered a member of Sgr~OB1. This is a very large OB association, that extends between $l = 4\degr$ and $l = 14\degr$ on both sides of the Galactic Plane, corresponding to a size of $\sim 300 \times 120$~pc, at an estimated distance of 1.7~kpc \citep{hum1978}. Nevertheless, Sgr~OB1 does not seem to be a homogeneous grouping. \citet{melef95} identified different spatial and kinematic subgroups. There is also a sizeable age spread. The open clusters NGC~6530 and NGC~6514 (M~20) are embedded in bright nebulosity (M~8 in the case of NGC~6530) and still forming stars, while the nearby NGC~6531 (M~21) has an estimated age of 12~Ma \citep{msg05}. \textit{Gaia} DR2 data have confirmed this lack of homogeneity. The proper motions of NGC~6530 ($\sim$~$+1.3$~mas~a$^{-1}$, $\sim$~$-2.0$~mas~a$^{-1}$; \citealt{damiani19}, \citealt{kuhn19}) are quite different from those of NGC~6514 ($\sim$~$+0.4$~mas~a$^{-1}$, $\sim$~$-1.7$~mas~a$^{-1}$; \citealt{kuhn19}), which is less than $1\fdg5$ away in the sky and has about the same age. Conversely, the proper motions of NGC~6514 are very similar to those of the much older NGC~6531 ($\sim$~$+0.5$~mas~a$^{-1}$, $\sim$~$-1.5$~mas~a$^{-1}$; \citealt{cantat18}).\\

The distances to these clusters, believed to be the core of the association, are rather lower than assumed forty years ago. \citet{damiani19} measure 1325~pc for NGC~6530, with a 9\,\% error. \citet{kuhn19} obtain $1336^{+76}_{-66}$~pc for NGC~6530 and $1264^{+78}_{-66}$~pc for NGC~6514. No specific study has been conducted for NGC~6531, but simple inversion of the average parallax found by \citet{cantat18} gives 1250~pc. All these values seem to agree on a distance $\sim1.3$~kpc for the association. VX~Sgr is located about one degree to the northeast of NGC~6531, the oldest cluster in the association. The proper motions derived by \citet{xu2018}, $\mu_{\alpha}\:\cos{\delta}$ = $0.36\pm0.76$~mas\,a$^{-1}$, $\mu_{\delta}$= $-2.92\pm0.78$~mas\,a$^{-1}$ may be considered consistent with those of the cluster. There are very few studies aimed at NGC~6531. Using Str\"{o}mgren photometry, \citet{msg05} derive an age of 12~Ma. This is roughly consistent with the value of $\sim8$~Ma found by \citet{forbes96}, as he was using earlier isochrones without overshooting, which result in younger ages. The brightest stars in the cluster are B1\,V, with the exception of HD~164863, which is slightly earlier \citep{forbes96}. With a mean $E(B-V)=0.28$, \citet{forbes96} finds a distance of $1.38\pm0.08$~kpc, consistent with the \textit{Gaia} values for the association.\\

In all, the very likely membership of VX~Sgr in Sgr~OB1 suggests that its distance should be close to 1.3~kpc, which is marginally in tension with the values derived from H$_{2}$O masers, although not fully inconsistent, and lower than our weighted average value of 1.44~kpc -- although in agreement with its GDR2 parallax. Assuming a distance of only 1.3~kpc for VX~Sgr would reduce its bolometric luminosity to $-8.4$~mag, still brighter than any known AGB star, but now roughly compatible within errors with theoretical values.\\

On the other hand, membership offers little constraint on its mass. It is obvious that an evolved red star is not directly connected to any of the star-forming clusters. Claiming a connection to NGC~6531 would imply a mass of $12$\,--\,$14\:\mathrm{M}_{\sun}$ for standard evolution and $\la25\:\mathrm{M}_{\sun}$ for the product of a merger (as B1\,V stars have around $10\:\mathrm{M}_{\sun}$; \citealt{harmanec88}). Although the halo of NGC~6531 does not extend to the location of VX~Sgr, the published spectral types of the early-type stars in its immediate vicinity, HD~165857 (B2\,III), HD~165689 (B2\,IV) or HD~165595 (B5\,III), do not suggest an earlier age. Conversely, no known star in the neighbourhood of VX~Sgr is suggestive of a massive ($M_*\ga20\:\mathrm{M}_{\sun}$) population, while its interpretation as a RSG requires a mass close to $40\:\mathrm{M}_{\sun}$ \citep{arr2015}.

\section{Spectral features}
\label{sp_feat}

\subsection{Spectral classification}
\label{spec_class}

We provided two different sets of SpTs, determined by using classification in the optical range ($4\,800$ to $7\,700\:$\AA{}) and  in the near infrared ($7\,800$ to $8\,900\:$\AA{}). Firstly, we performed the optical classification using the  procedure detailed in \citet[][]{dor2018b} to derive the SpTs from the SES spectra. Unfortunately, the spectra from UVES do not cover any wavelength bluer than $6\,750\:$\AA{} and, in consequence, we employed only the TiO band systems located at $7\,050\:$\AA{} and at $7\,600\:$\AA{}. Next, we performed the near-infrared classification according to the criteria proposed by \cite{sha1956} and \cite{sol1978}. To that aim, we employed the atlases published by \cite{gin1994} and \cite{car1997} following the classification procedure presented in \cite{neg2011,neg2012} in the Calcium Triplet range ($8\,500\:$\AA{} to $8\,900\:$\AA{}).\\

We present the calculated SpTs in Table~\ref{epochs}, all of which are accurate down to one spectral subtype. These two classification methods lead to divergent SpTs, depending on the wavelength region  considered. Our spectral types in the near infrared are later than those obtained by using the optical range \citep[this effect had already been reported for VX~Sgr by][]{hum1974a}. Only the spectrum observed on 2018-06-06 is compatible in both spectral ranges, as it is M3 in the optical and M4 in the near-infrared.\\

\begin{figure}
   \centering
   \includegraphics[width=\columnwidth]{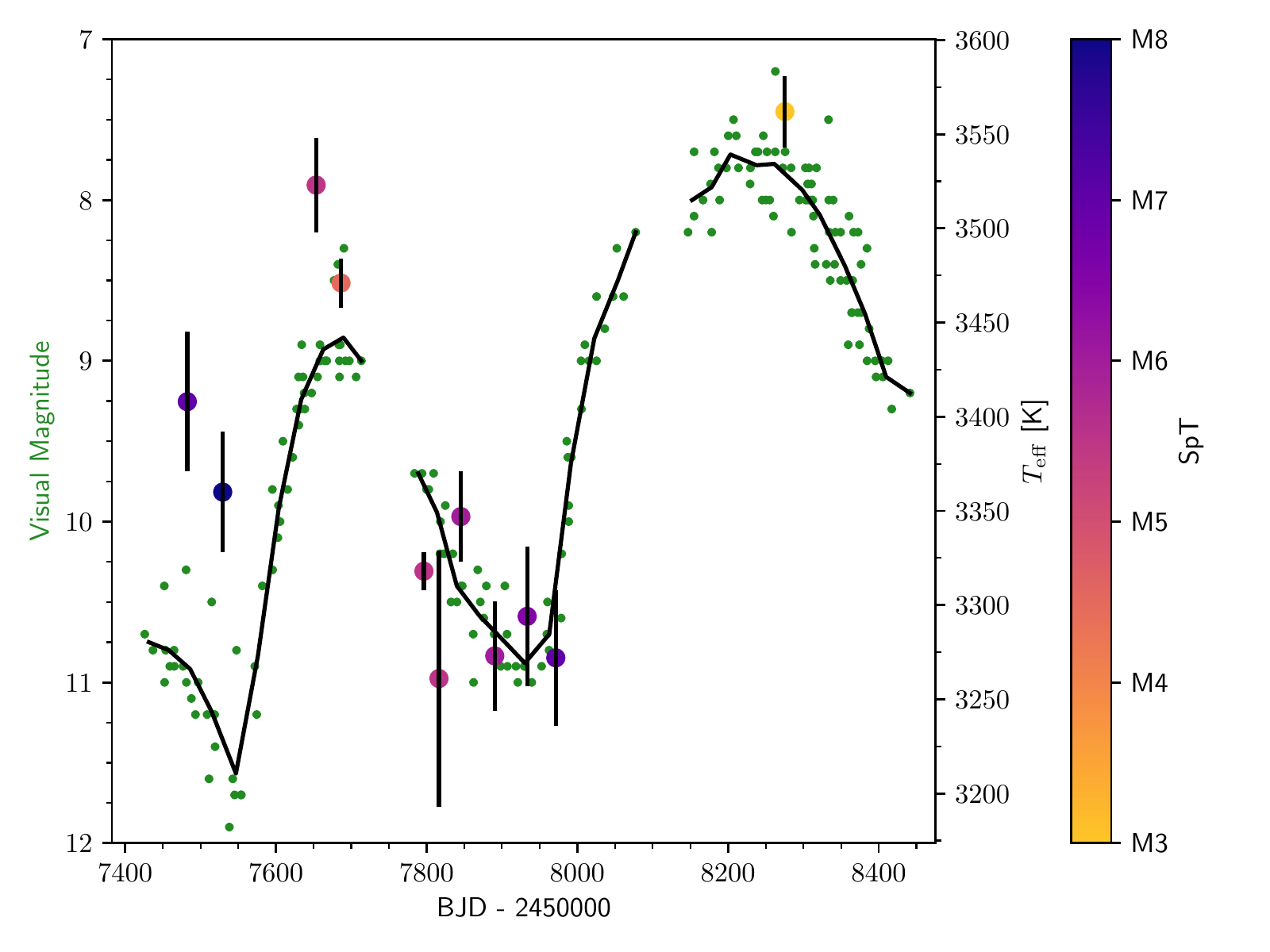}
   \caption{Same as Fig.~\ref{recent_rv}, but now circles indicate the $T_{\rm eff}$ of the corresponding spectrum (right-y axis) while their colour indicates the optical SpT classification.} 
   \label{recent_teff}
\end{figure}

 As spectral type and temperature show a large degree of correlation for earlier types \citep[up to M3; see ][and references therein]{tab18}, we checked for a correlation between $T_{\rm eff}$ and SpT for both the optical and near-infrared spectral types derived in this work. We calculated a correlation of  $r=-0.74\pm0.22$ for the infrared classification and $r=-0.59\pm0.27$ for the optical. Thus, there is a correlation between the SpTs and $T_{\mathrm{eff}}$ regardless of the spectral classification criteria employed, albeit the correlation is weaker in the case of the optical classification. In addition, we found that our effective temperatures follow the variations of the light curve (see Fig~\ref{recent_teff}). In consequence, we would expect the SpTs to follow the variations of the light curve to some extent, although VX~Sgr changes over a wide range of SpTs that are not well connected to $T_{\mathrm{eff}}$ by any modern temperature scale. In any case, these results support the idea that the light variation is linked to changes in the opacity of molecular layers in the upper part of the atmosphere \citep{hum1974a,kam2005}, caused by changes in the $T_{\mathrm{eff}}$.\\

\subsection{Atomic features}
\label{atomic_features}
 
\begin{figure*}         
    \includegraphics[trim=0.8cm 0.2cm 0.5cm 1cm,clip,width=9cm]{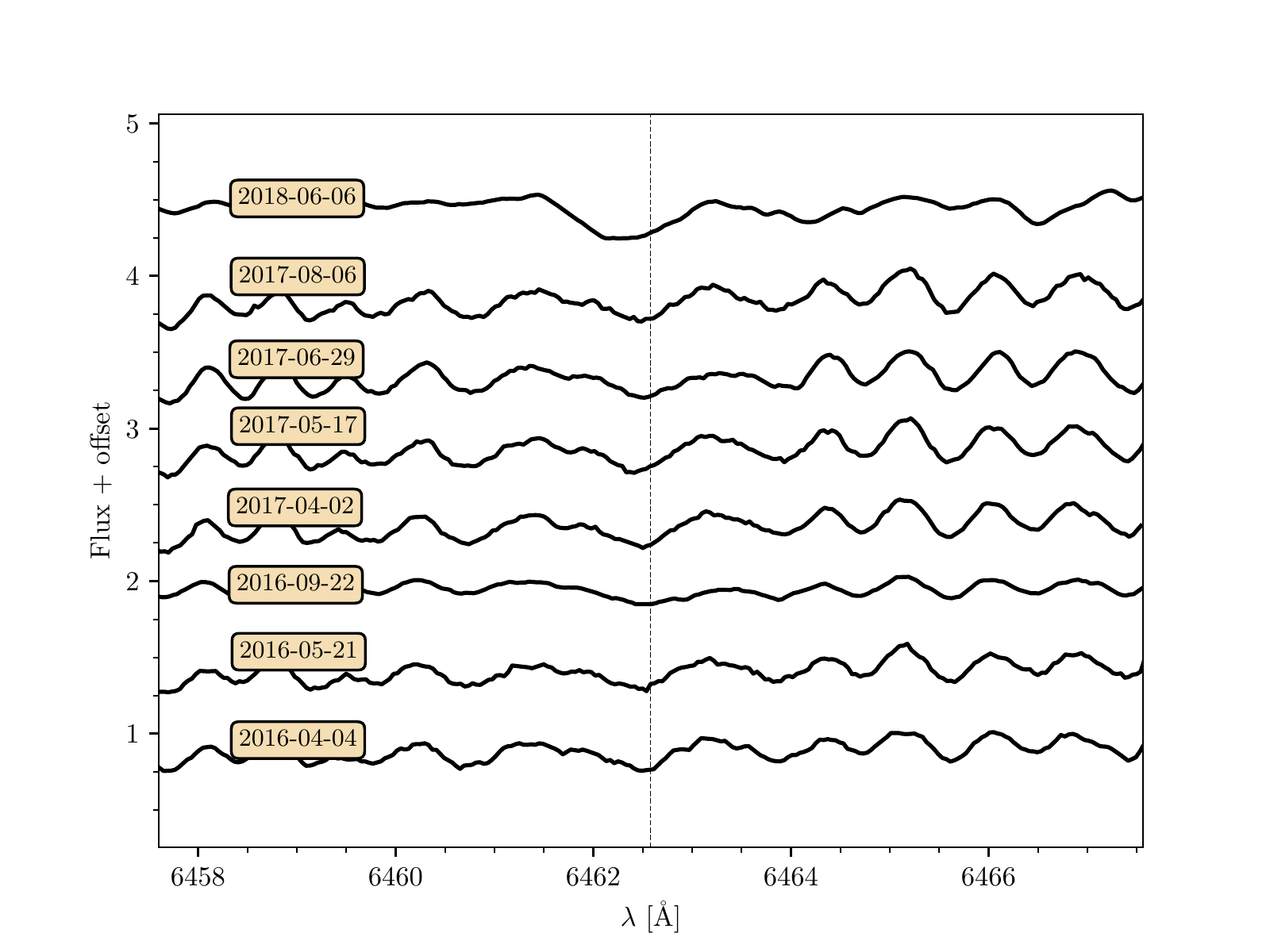}
    \includegraphics[trim=0.8cm 0.2cm 0.5cm 1cm,clip,width=9cm]{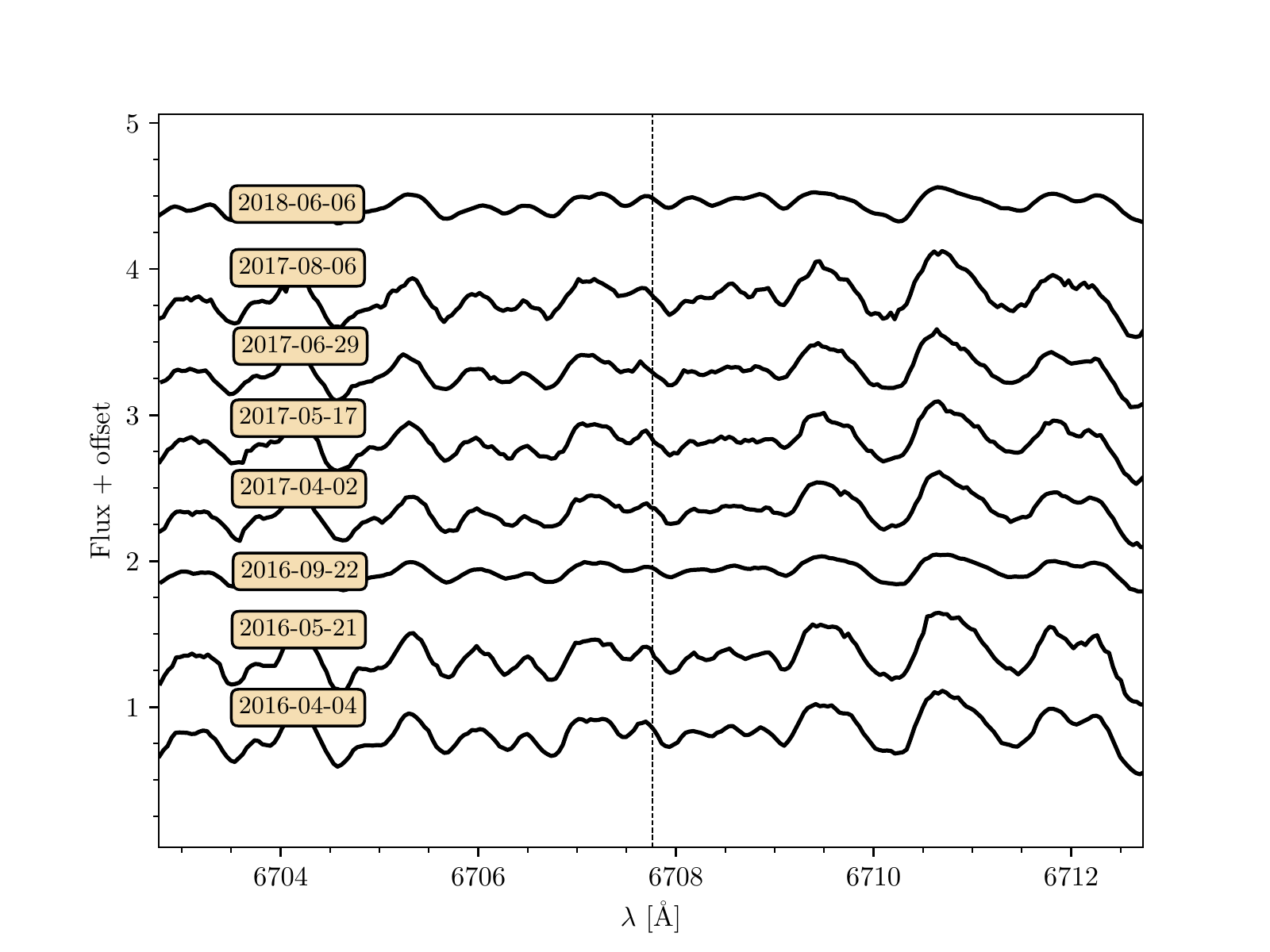}
    \caption{The Spectrum of VX~Sgr around the \ion{Ca}{i} line at 6462.58~\AA{} (left) and the \ion{Li}{i} line at 6707.76~\AA{} (right). The spectra have been tagged with the dates given in Table~1, whereas the centre of the line is indicated by a black vertical line.}
    \label{litio_calcio}                                     
\end{figure*}

We explored our spectra looking for the presence of \ion{Li}{i}, \ion{Ca}{i}, and \ion{Rb}{i}. The presence of these lines would be connected to internal processes at work inside VX~Sgr. Following \citet{per2019} we inspected the \ion{Li}{i} at 6707.76~\AA{} and the \ion{Ca}{i} line at 6462.58~\AA{}. We did not find any signs of \ion{Li}{i} and the \ion{Ca}{i} line is barely present in our observations (see Fig.~\ref{litio_calcio}).\\

\begin{figure*}                                               
    \includegraphics[trim=0.8cm 0.2cm 0.5cm 1cm,clip,width=9cm]{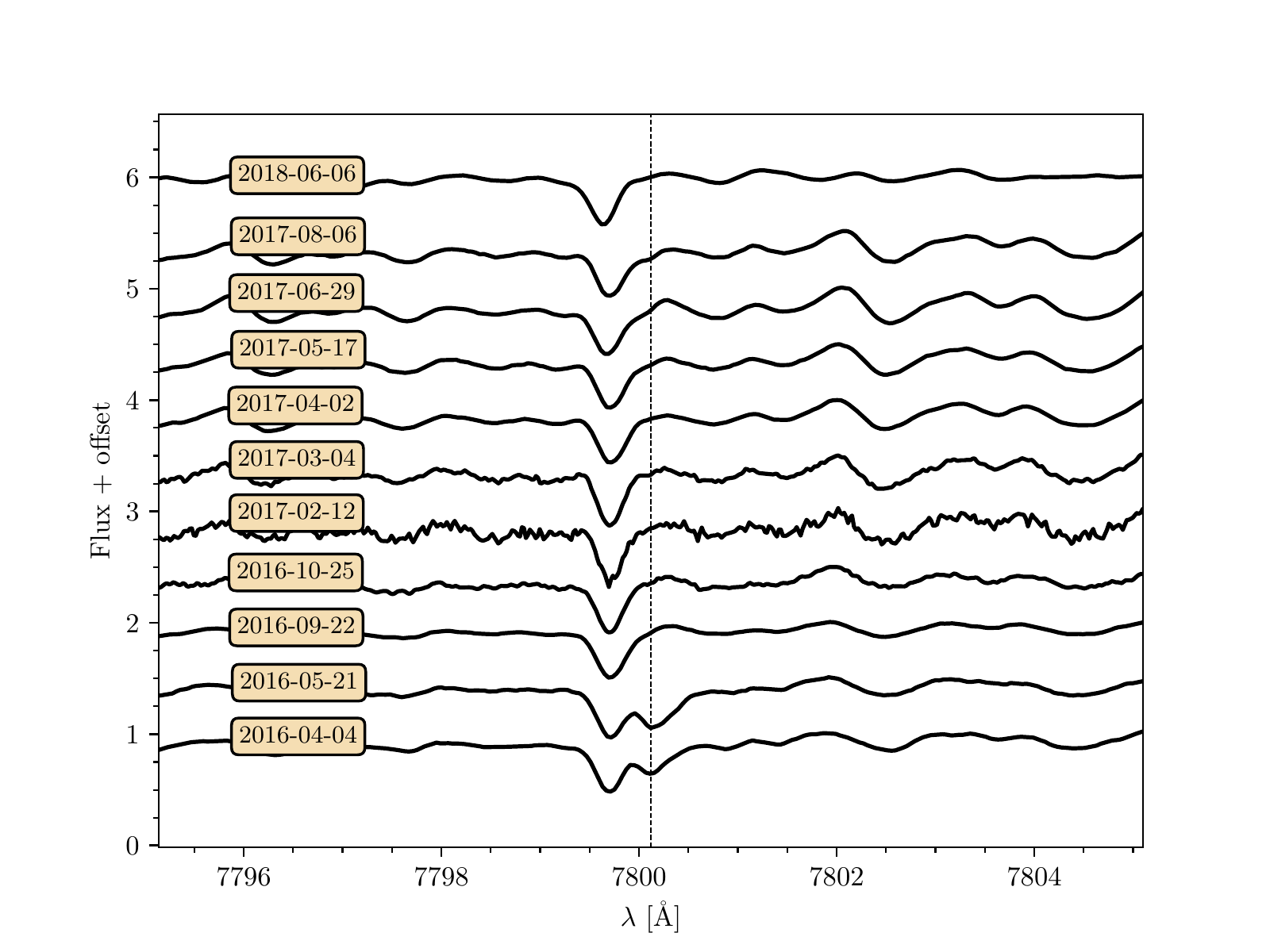}                  
    \includegraphics[trim=0.8cm 0.2cm 0.5cm 1cm,clip,width=9cm]{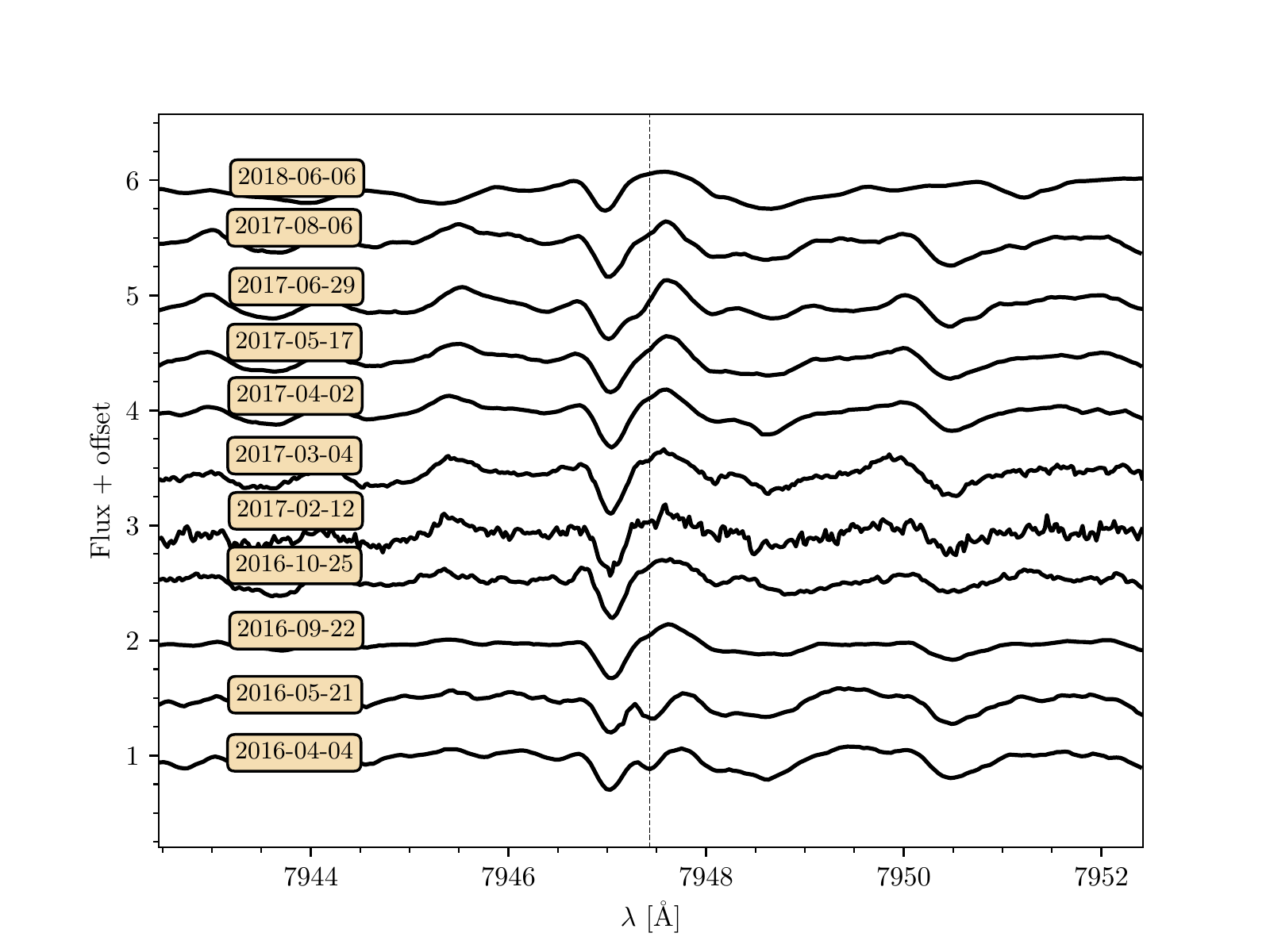}
    \caption{ Same as Fig.~\ref{litio_calcio} but for the \ion{Rb}{i} lines at $7\,800.26\:$\AA{} (left) and $7\,947.60\:$\AA{} (right).}
    \label{rubidium}                                         
\end{figure*}

\begin{figure*}                                               
    \includegraphics[trim=0.8cm 0.2cm 0.5cm 1cm,clip,width=9cm]{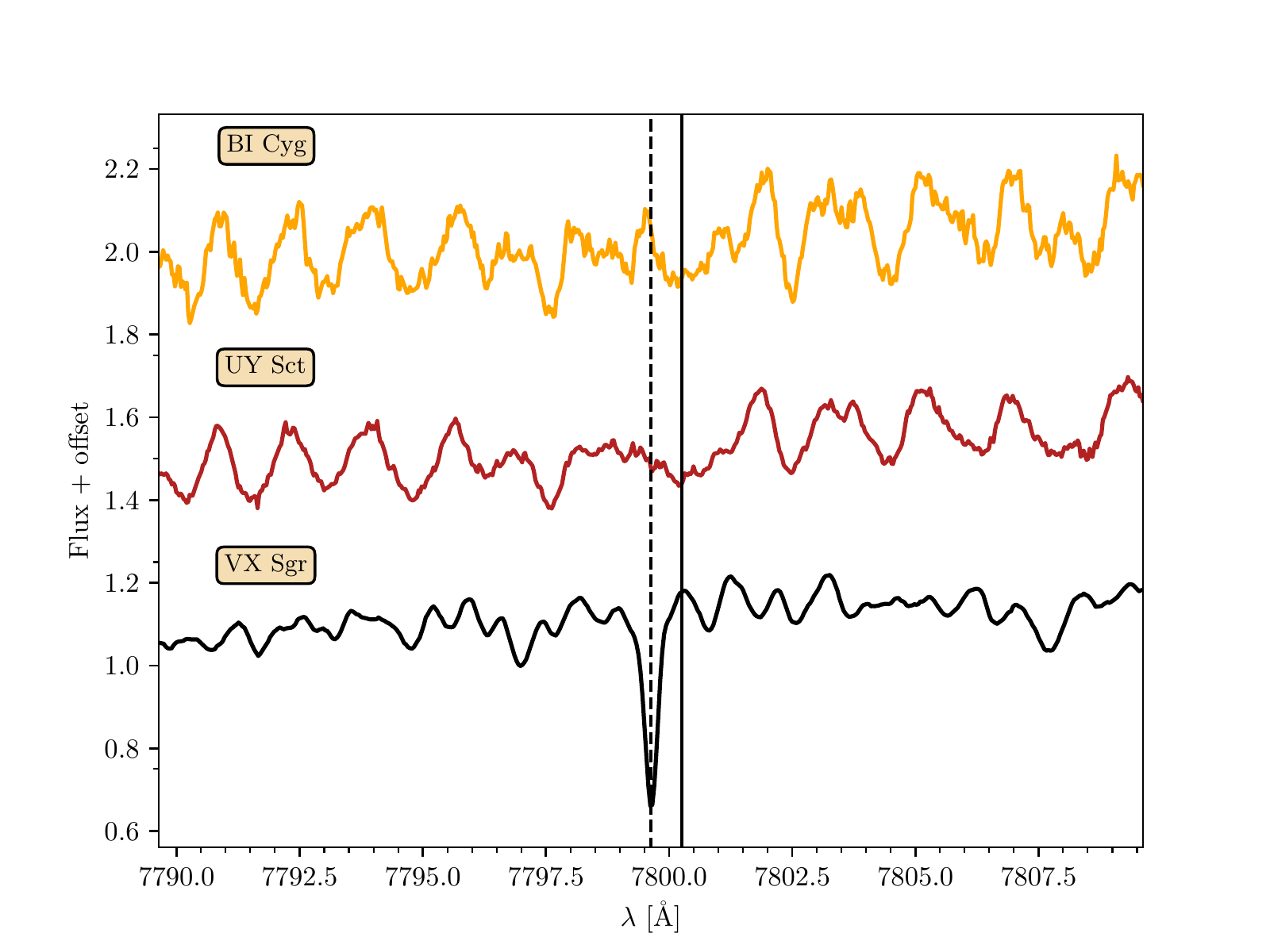}                  
    \includegraphics[trim=0.8cm 0.2cm 0.5cm 1cm,clip,width=9cm]{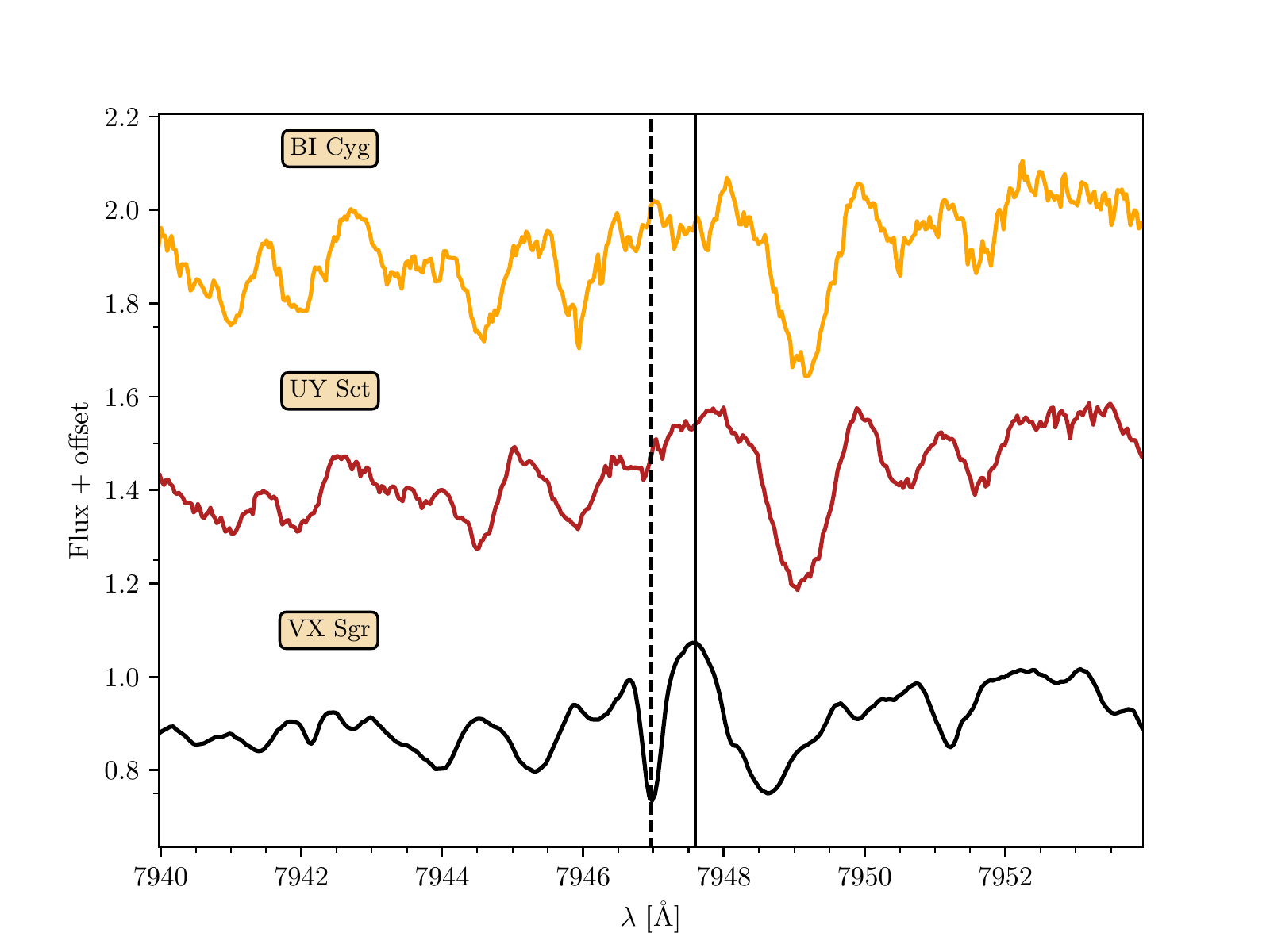}
    \caption{Comparison of the spectrum of VX~Sgr (2018-06-06, black) against the spectrum of UY~Sct amd BI~Cyg  observed with the SES spectrograph around the \ion{Rb}{i} lines at $7\,800.26\:$\AA{} (left) and $7\,947.60\:$\AA{} (right). The vertical line indicates the position of the \ion{Rb}{i} lines, and the dashed vertical line indicate the position of the circumstellar \ion{Rb}{i} component.}
    \label{rubidium_compare}                                         
\end{figure*}

 The presence of \ion{Rb}{i} in the spectrum of VX~Sgr was already reported by \citet{gar2006}. Consequently, we confirm that two Rubidium lines, namely  $7\,800.26\:$\AA{} and $7\,947.60\:$\AA{}, are visible in the spectrum of VX~Sgr (see Fig.~\ref{rubidium}). We found that these two lines were equally blueshifted with respect to their rest-frame wavelengths. We measured an average blueshift of $-21.5\pm0.9\:$km~s$^{-1}$ for both \ion{Rb}{i} lines (see Table~\ref{rubidium_shift}), which is consistent with the expansion velocities derived from OH masers already reported in the literature \citep[19.8 and 23.0 km~s$^{-1}$ in ][respectively]{lin1989,che2007}. This agreement suggests that these blueshifted profiles are circumstellar, which is also the interpretation given by \citet{gar2006}. Unfortunately, \citet{gar2006} did not provide any value for the Rb abundance of VX~Sgr.  However, the circumstellar component of VX~Sgr is one of the strongest in their sample, which suggests a fair amount of Rb enhancement. To visualize it, we compare the \ion{Rb}{i} lines of VX~Sgr to those of the well known RHGs BI~Cyg and UY~Sct  \citep[see ][]{hum1983,arr2013} in Fig.~\ref{rubidium_compare}. The plot shows that no \ion{Rb}{i} lines (either photospheric or circumstellar) are present in the spectra of UY~Sct and BI~Cyg.\\ 

Interestingly, two of our spectra (2016-04-04 and 2016-05-21, see Fig.~\ref{rubidium}) showed both photospheric and circumstellar components of the \ion{Rb}{i} lines. According to the radiative transfer simulations of \citet{zam2014}, their shape is connected to changes in the density of the circumstellar medium with respect to other epochs. A strong P-Cygni profile develops, hiding the photospheric companion and resulting in the observed blueshift. Given that these two spectra are the first we acquired, we cannot know for how long these double features were present. However, we know that in the next spectrum, taken four months later, the photospheric Rb component is missing. In fact, there is no trace of the photospheric \ion{Rb}{i} lines in any of our spectra after these two, although they cover most of the magnitude range along the AAVSO light curve.\\

Finally, we examined the Balmer lines in the spectra for any signs of stellar activity. We found that Balmer lines are in emission on the two epochs close to the maximum brightness (see Table~\ref{epochs}). A close-up view  of the spectra around the Balmer lines up to H$\delta$ is shown in Fig.~\ref{Balmer}. We note that emission lines at light maxima had already been reported by \citet{hum1974a}.

\begin{table}
\tiny
\caption{Measured line centres of the circumstellar \ion{Rb}{i} features.}
\label{rubidium_shift}
\centering
\begin{tabular}{ccccc}
\hline
&  \multicolumn{2}{c}{Rb~7800.26~\AA} & \multicolumn{2}{c}{Rb~7947.60~\AA}   \\
Date &  Centroid &  V$_{\rm shift}$ & Centroid & V$_{\rm shift}$ \\
    &  [\AA]  & [km~s$^{-1}$] &  [\AA] & [km~s$^{-1}$]  \\
\hline 
\hline
\noalign{\smallskip}
2016-04-04 & 7799.71 & -21.1 & 7947.03 & -21.4  \\
2016-05-21 & 7799.72 & -20.7 & 7947.04 & -21.0 \\
2016-09-22 & 7799.69 & -21.9 & 7947.05 & -20.6  \\
2016-10-25 & 7799.71 & -21.1 & 7947.06 & -20.3 \\
2017-02-12 & 7799.70 & -21.5 & 7947.03 & -21.4 \\
2017-03-04 & 7799.70 & -21.5 & 7947.03 & -21.4 \\
2017-04-02 & 7799.72 & -20.7 & 7947.05 & -20.6 \\
2017-05-17 & 7799.71 & -21.1 & 7947.04 & -21.0 \\
2017-06-29 & 7799.71 & -21.1 & 7947.03 & -21.4 \\
2017-08-06 & 7799.69 & -21.9 & 7947.01 & -22.2 \\
2018-06-06 & 7799.63 & -24.2 & 7946.97 & -23.7 \\
 \hline
\end{tabular}
\end{table}

\subsection{Line-doubling}
\label{duplicated}
   
\begin{figure}         
    \includegraphics[width=\columnwidth]{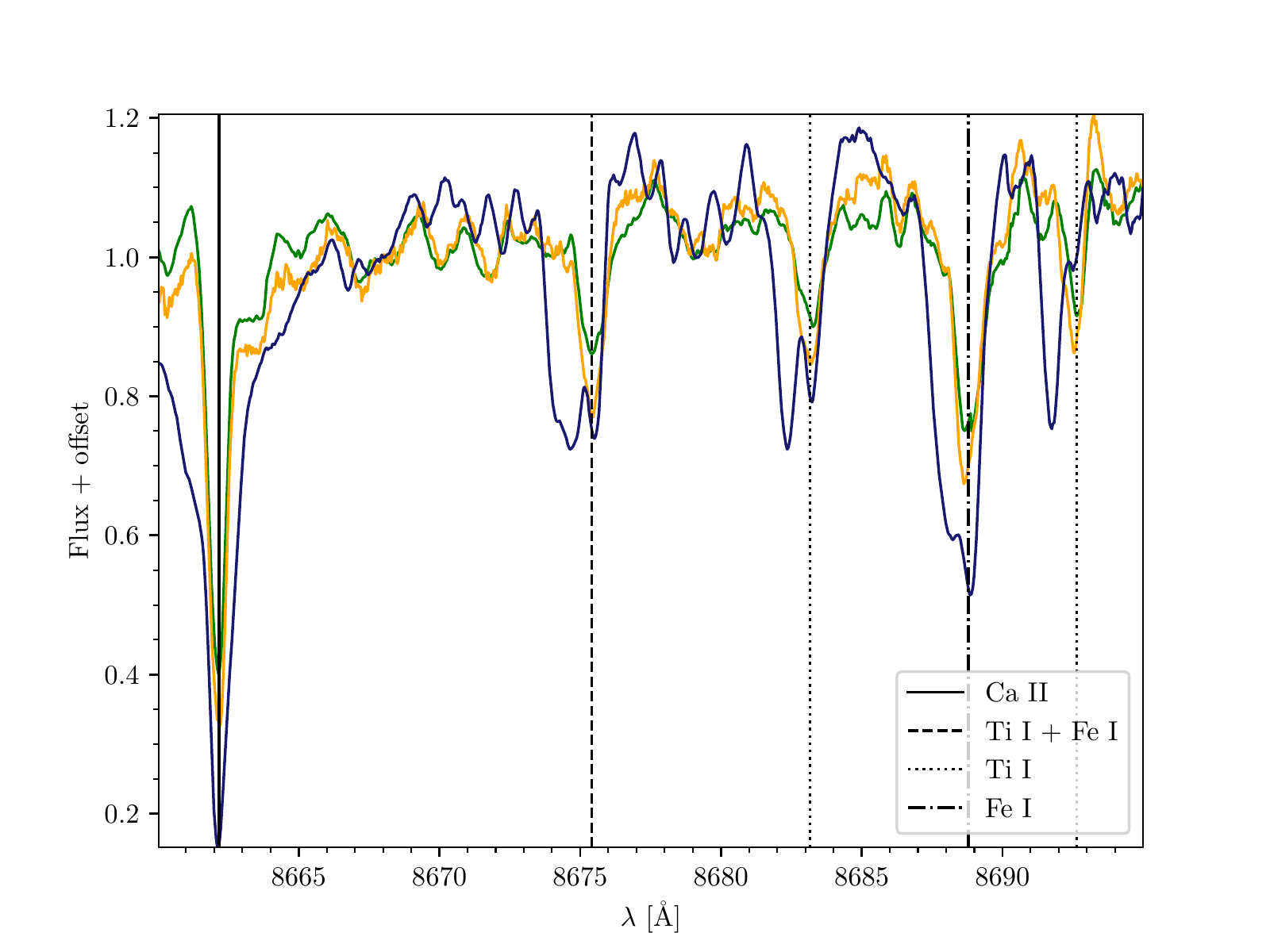}
    \caption{Example of the line doubling of atomic features observed on the 2018-06-06 spectrum (green). For comparison, two other spectra obtained during the first light maximum (2016-09-22 in blue, and 2016-10-25 in orange) are shown.}
    \label{line_dup}                                        
\end{figure}

In the spectrum obtained during the second maximum light (2018-06-06, see Fig.~\ref{line_dup}), the last in our series, most atomic lines (mainly those of \ion{Ti}{i} and \ion{Fe}{i}) appear double. This effect is not observed on any other epoch. Line doubling effects are well reported for all types of luminous cool stars \citep{alv2001}, and believed to be caused by convective movements in the stellar atmosphere. This type of line doubling is unrelated to the presence of the circumstellar components observed in the \ion{Rb}{i}~lines.\\

Line doubling has been observed in both RSGs and Mira stars, but its cause is different in each case, resulting in its appearance at different phases of the light curve \citep{jor2016}. In Mira stars, this effect is caused by the ascending movement of a shock wave through the atmosphere, in what is known as ``Schwarzschild scenario''. This shock wave is connected to the stellar pulsation, and thus, line-doubling is seen around the maximum light of the photometric curve. On the contrary, in RSGs the duplication is caused by the rising of a new large convective cell \citep{kra2019}. In consequence, this effect, when observed, takes place during the ascending part or the light curve.\\

\section{Period analysis}
\label{period}

\begin{figure}
   \centering
   \includegraphics[trim=0.2cm 0.2cm 0.5cm 1cm,clip,width=9cm]{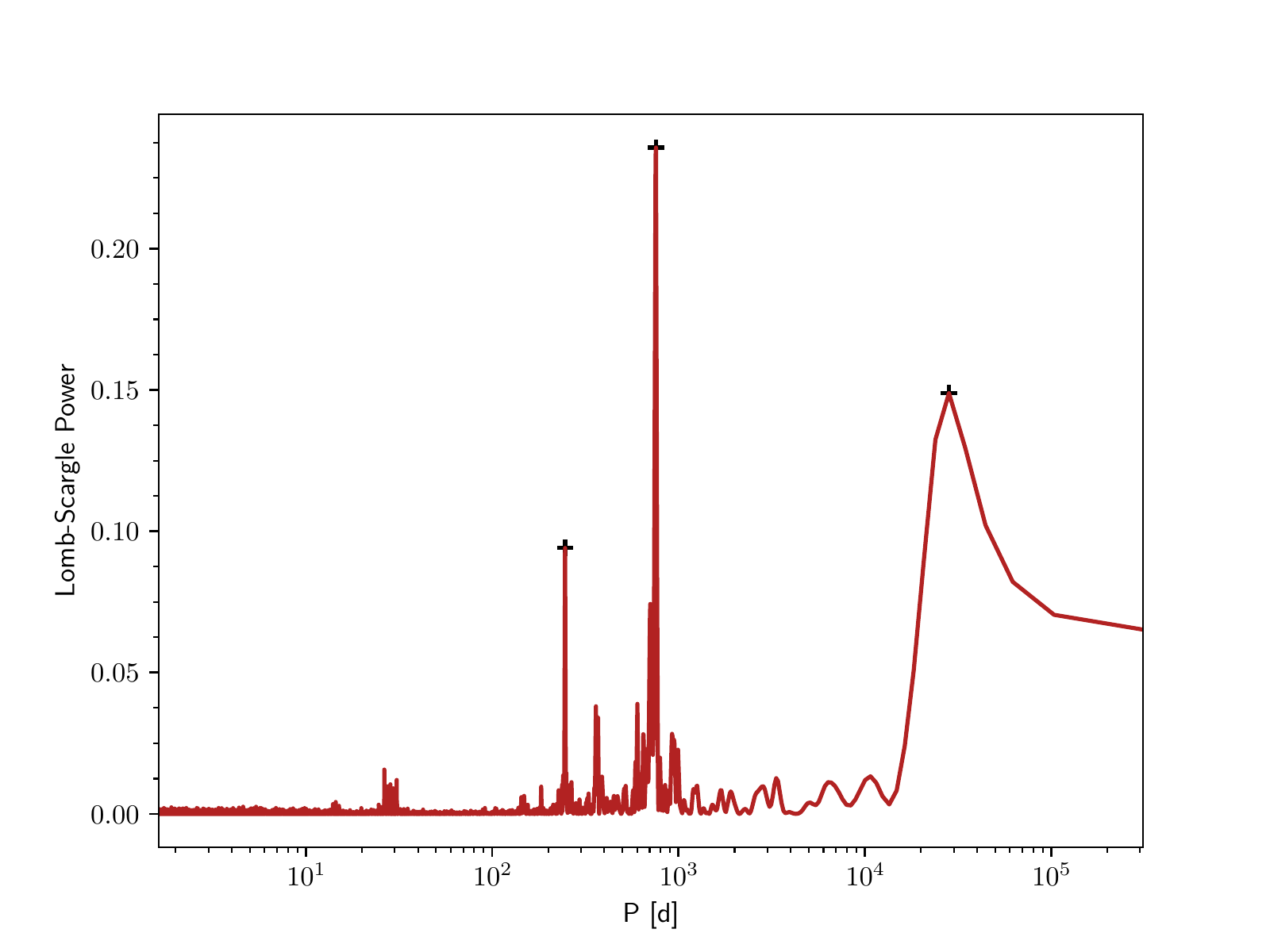}
   \caption{Lomb-Scargle periodogram generated using the AAVSO light curve. The black crosses show the three signals found at 246~d, 757~d and $28\,279\:$~d.}
   \label{periodogram}
\end{figure}

\begin{figure*}
   \centering
   \includegraphics[trim=0.2cm 0.2cm 0.5cm 1cm,clip,width=\textwidth]{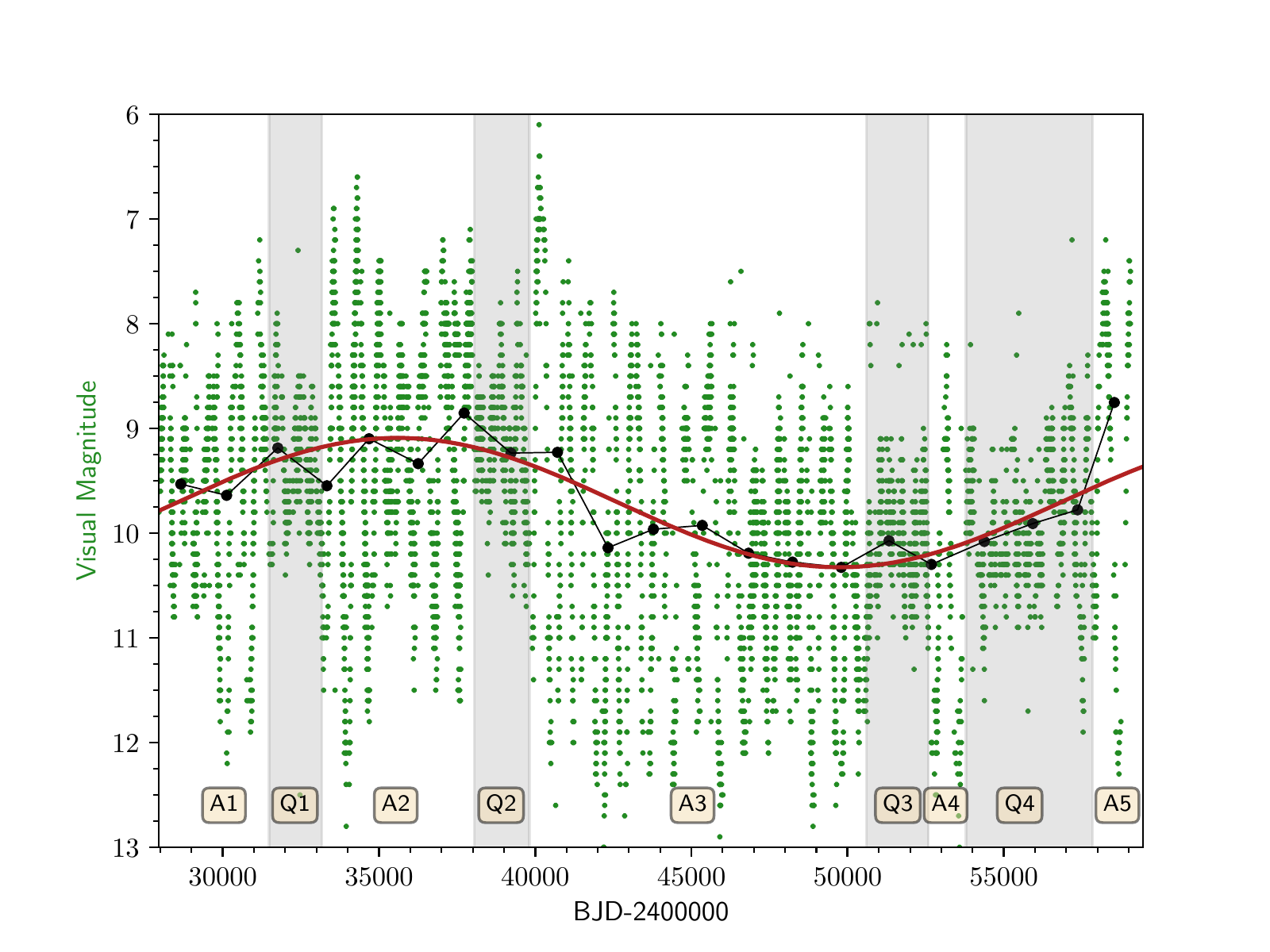}
   \caption{Light curve of VX Sgr, from 1935-05-25 (BJD~2427948) to 2020-07-27 (BJD~2459055). Green dots indicate the visual magnitudes that make up the light curve. The black dots represent a $1\,514\:$d binning of the light curve (i.e. two periods of 757\:d). The red curve is a sinusoidal toy model with a $28\,279\:$d ($\sim77\:$a) period. The quiescent phases are shaded in grey and indicated at the bottom (see Table~\ref{phases}).}
   \label{lightcurve}
\end{figure*}

The light curve of VX~Sgr has been continuously registered by the AAVSO since 1935. This allowed \cite{kuk1969} first, and \cite{kam2005} later, to calculate the main photometric period of VX~Sgr at $732\:$ d and $\sim750\:$ d, respectively. We calculated the main period again through the Lomb-Scargle periodogram (see Figure~\ref{periodogram}). We obtained a period at 757\:d, which is similar to that obtained by \cite{kam2005}.  Interestingly enough, our data show a peak at $28\,279\:$d ($\sim77\:$a) that has never been reported before. Previous studies, such as \cite{kuk1969} lacked enough data coverage, while \cite{kam2005} dismissed periods above $12\,000\:$d.  Suggestively, a sinusoidal with this long period is in good agreement with the smoothed light curve shown in Figure~\ref{lightcurve}. However, this putative period at $28\,279\:$d has roughly the same time-span as the whole AAVSO lightcurve for VX~Sgr, while at least two complete periods are required to confirm it. Finally, the third period found at 246\:d is spurious ($757^{-1}+365^{-1}\approx246^{-1}\:$d$^{-1}$) and should be discarded altogether.\\

Although the periodogram of VX~Sgr exhibits a well-defined main period of 757~d, its light curve shows shows two different behaviours that have not been previously studied in detail. VX~Sgr sometimes exhibits large photometric variability with peak-to-peak amplitudes between 4 and 6~mag, whereas at another times the peak-to-peak amplitudes are at about 2~mag. We refer to the first as active phases, and the second as quiescent phases. To study these two different photometric behaviours, we partitioned the light curve into 9 different phases, 4 characterised by quiescence and 5 by activity. These phases are shown in Fig.~\ref{lightcurve} and described in Table~\ref{phases}. We calculated individual periodograms for each of the active (A1-A5) and quiescent phases (Q1-Q4) defined. We found a period of $\sim770\:$d for phases A1, A3, and A4, whereas active phase A2 gave a shorter period of 720\:d. In contrast, the current active phase (A5) displays a longer period of $\sim$931\:d. However, this value must be considered with caution, as the present phase (A5) has not spanned two full periods yet and, as the phase progresses, the final period may result in a different value. Moreover, the start of the current phase is not well defined. The faintest minimum since the A4 and the maximum observed in 2016 have an appearance typical of active phases, but the variations are too fast compared with active phases, and so we left them out of A5. The time elapsed between the 2016 minimum and the next one, in 2017, is only $\sim390\:$d, while the time between the 2016 maximum until the 2018 one is only $\sim500\:$d. These values are more typical of the variations seen during quiescent phases, an fit well with the rest of the phase Q4.\\

The quiescent phases reveal a different periodic behaviour. Not only the amplitude of the light curve decreases significantly, but also the well defined periodicity seems to disappear. We obtained periods at 583, 652, and 624\;d for quiescent phases Q1, Q2, and Q4 respectively, whereas Q3 revealed three roughly equally tall peaks at 239, 685, and 1527\:d. The signal at 1527\:d has to be taken with care because Q3 does not cover the time span necessary to fully sample it, whereas, the 239\:d signal  is spurious ($685^{-1}+365^{-1}\approx239^{-1}\:$d$^{-1}$) and should be discarded. On the light of these results, we investigated the periodogram for the whole historical record of VX~Sgr (see Fig.~\ref{periodogram}) and found a weak peak at roughly $\sim602\:$d, which, in turn, is similar to the periodic signals that we found for quiescent phases. 

\begin{table*}
\centering
\begin{tabular}{c  c c  c c  c c }
\hline
& \multicolumn{2}{c}{Beginning}  & \multicolumn{2}{c}{End} &   &  \\
Phase       & BJD     & Date        & BJD    & Date        &  Time span  &  Period      \\
            &         &             &        &             &  [d]        & [d] \\
\hline
\hline
\noalign{\smallskip}
A1    & 2427948 & 1935-05-25 & 2431440 & 1944-12-15 & 3492    & 770\\
Q1  & 2431440 & 1944-12-15 & 2433170 & 1949-09-10 & 1730    & 583  \\
A2    & 2433170 & 1949-09-10 & 2438030 & 1962-12-31 & 4860   & 720      \\
Q2 & 2438030 & 1962-12-31 & 2439860 & 1968-01-04  & 1830   & 652 \\
A3    & 2439860 & 1968-01-04 & 2450600 & 1997-05-31 & 10740  & 770      \\
Q3 & 2450600 & 1997-05-31 & 2452650 & 2003-01-10 & 2050   & 685 \\
A4    & 2452650 & 2003-01-10 & 2453740 & 2006-01-04  & 1090    & 768  \\
Q4 & 2453740 & 2006-01-04 & 2457845 & 2017-04-01  & 4105   & 624 \\
A5    & 2457845 & 2017-04-01 & 2459055 & 2020-07-27  & 1210   & 931 \\
\hline
\end{tabular}
\caption{  Partition of the light curve of VX~Sgr into five active (i.e. A1-A5) and four quiescent (i.e. Q1-Q4) phases. All periods were calculated through the Lomb-Scargle periodogram.}
\label{phases}
\end{table*}

\section{Discussion}
\label{discussion}

\subsection{The nature of VX~Sgr}
\label{nature}
 VX~Sgr is a fundamentally challenging star, given its characteristics. We consider first the possibility that VX~Sgr might be a T\.ZO, because these hypothetical stars can theoretically reach the luminosities of RSGs and they can produce a distintict chemical signature in their spectra.  However, the \ion{Li}{i} at 6707.76~\AA{} line is not seen and the \ion{Ca}{i} line at 6462.58~\AA{} is barely present in our observations (see Fig.~\ref{litio_calcio}), while they are expected in a T\.ZO. Thus, we should rule out the T\.ZO scenario in favor of other possibilities such as VX~Sgr being either a RSG/RHG or an AGB.\\

The key feature to understand VX~Sgr is the presence of strong \ion{Rb}{i} lines in its spectrum, indicative of Rb overabundance  when compared to other RHGs as shown in Fig.~\ref{rubidium_compare} (BI~Cyg and UY~Sct, see Sect.~\ref{atomic_features}). These \ion{Rb}{i} lines are regularly detected in the spectra of candidate high-mass AGB stars \citep[those with masses $\geq3M_{\odot}$ ][]{gar2006,gar2007}.  However, an enhanced Rb, as well as any other s-element, is completely unexpected in RSGs or RHGs, as there are not theoretical mechanisms capable of producing a Rb overabundance in the atmosphere of these stars. In fact, these lines have not been observed in RSGs \citep[e.g.][see also Fig.~\ref{rubidium_compare}]{smi1995,kuc2002}. Thus, we should rule out the RSG/RHG scenario.\\

The Rb overabundance in AGB stars is a direct consequence of the s-process taking place in the stellar interior. Rb is dragged out to the surface later, during the third dredge-up that happens between thermal pulses \citep[TPs;][]{her2005}. The efficient production of Rb only takes place in 
massive AGB stars \citep[$\gtrsim$4\,--\,5~M$_{\odot}$, see][]{gar2006,kar2012,raa2012}. In parallel, the Cameron-Fowler mechanism \citep[][]{cam1971, maz1999} also happens in the stellar interior, and results in a lithium enrichment of the surface. However, while the enrichment of s-process elements (such as Rb) increases slowly over time, Li production is much faster, reaching a maximum early in the TP phase and then slowly decaying \citep{maz1999, raa2012}. The presence of Rb together with the absence of Li (as is the case of VX Sgr) is expected in AGB stars at the end of their TP phase. Alternatively, this situation can happen, according to models \citep{raa2012,doh2014a}, in some AGB stars earlier in the TP phase, due to temporary drops in Li abundance. However, Li drops are predicted to be smaller at higher stellar masses. An example of this can be seen in \citep{gar2013}. Thus, a drop that results in non-detection of Li can only happen in intermediate-mass AGB star (between $\sim4$\,--\,7\:M$_{\odot}$).\\

The high luminosity of VX~Sgr is simply incompatible with such an intermediate mass ($\sim4$\,--\,7\:M$_{\odot}$). In addition, a number of facts hint at the idea that VX~Sgr has depleted its Li because it is close to the end of its TP phase. Firstly, the abundance of Rb grows slowly over time, and does not  experience significant drops during the TP phase. In consequence, the higher the abundance, the more evolved the AGB star must be. Although we cannot obtain a precise measurement of the Rb abundance, there is a strong reason to believe that the abundance of Rb is very high: \citet{gar2006} found that the Rb abundance is strongly correlated to the expansion velocity of the circumstellar envelope derived from OH masers ($v_{\mathrm{exp}}(\mathrm{OH})$). All the AGB stars with measured [Rb/Fe] in the sample of \citet{gar2006} have $v_{\mathrm{exp}}(\mathrm{OH})<16\:$km~s$^{-1}$. Those with the highest velocities have [Rb/Fe] values slightly below $2.5\:$dex. In the case of VX~Sgr its $v_{\mathrm{exp}}(\mathrm{OH})$ is even higher: \cite{lin1989} measured a $v_{\mathrm {exp}}(\textrm{OH})=19.8\:$km~s$^{-1}$, while \citet{che2007} obtained $v_{\mathrm {exp}}(\textrm{OH})=23.0\:$km~s$^{-1}$. These extreme values suggest that the Rb abundance of VX~Sgr is higher than any other in the sample of \citet{gar2006}. Secondly, according to \citep{gar2007}, $v_{\mathrm {exp}}(\textrm{OH})$ is also correlated to the Li abundance. However, two of the three stars in their sample with $v_{\mathrm{exp}}(\mathrm{OH})>16\:$km~s$^{-1}$ have no Li (including VX Sgr), which suggests that they are evolved enough to have depleted their Li. Thirdly, the pulsation periods of AGB stars increase during the TP phase \citep{vas1993,doh2014a}.  In fact, VX~Sgr has the longest period in the stellar sample studied by \citet{gar2007}, and it is the only star among them having a pulsation period above 500\:d that has no Lithium, all which is indicative of a more evolved stage. Thus, the presence of a significant circumstellar Rb strong absorption  \citep{gar2006} suggests that VX Sgr had time enough since the TP phase started to accumulate such high abundance of Rb in its circumstellar envelope.\\

At this point, it seems that VX~Sgr is a massive AGB star close to the end of its TP phase. Nevertheless, there is a major objection to this scenario. In comparison with the most extreme TP~AGB stars, VX~Sgr seems to have a short main photometric period and a low circumstellar reddening. Moreover, these behaviours are linked. Theoretical models \citep{vas1993,doh2014a} predict that, during the TP phase, the photometric period becomes longer. \cite{vas1993} proposed that, once the AGB reaches $P=500\:$d, the mass-loss rate undergoes a sudden increase, that according to the most recent evidence can reach $3\times10^{-5}\:\mathrm{M}_{\odot}\,\mathrm{a}^{-1}$ \citep{dec2019}. This effect is known as superwind and should take place at some point during the TP phase. After the onset of the superwind, the photometric period grows very quickly (up to 2000~d). Observations show that the Rb-rich AGB stars with the longest periods (500\,--\,1500\:d) are also those with the largest circumstellar reddenings and all of them have mass-loss rates around $10^{-5}\:\mathrm{M}_{\odot}\,\mathrm{a}^{-1}$ \citep{gar2006,per2017}.\\

A partial answer to this objection may be found in the work of \cite{kar2012}. The abundance of Rb and the ratio [Zr/Rb] obtained by theoretical models are significantly lower than those measured in observations \citep{per2017}. The reason for this discrepancy lies in a property of the theoretical models: the activation of the superwind causes a dramatic slowdown in the s-process enhancement \citep[see ][]{kar2012}. In order to obtain the s-process element enhancement needed to match the observations, \cite{kar2012} delayed the superwind phase until a period of 700\,--\,800\:d is reached. The abundances predicted by these superwind models (for intermediate-mass stars of 5, 6, and 7\:M$_{\odot}$) have been validated with observational data by \cite{gar2013}.\\

\cite{gar2007} showed that the IRAS colours of VX Sgr ($[12]-[25]=-0.74$ and $[25]-[60]=-1.8$) are similarly blue to those of AGB stars whose superwind has not started (those with short, $<400$\:d, and intermediate, $<700$\:d, periods). In contrast, the vast majority of AGB stars with long periods ($>700$\:d) exhibit much redder colours. Thus, VX~Sgr has a comparatively thin envelope. This supports the delayed superwind scenario as an explanation for the characteristics of VX~Sgr. Moreover, the lack of a thick circumstellar envelope in VX~Sgr despite its high mass-loss rate, 1\,--\,$6\times10^{-5}\:\mathrm{M}_{\odot}\,\mathrm{a}^{-1}$ \citep{kna1989,bec2010,mau2011,liu2017,gor2018}, suggests that its superwind phase may have started very recently. This is coherent with the typical delay in the superwind assumed by \cite{kar2012}.
\subsection{The mass of VX~Sgr}
\label{mass}

Although the nature of VX~Sgr, a massive AGB star close to the end of its TP-phase seems to be clear, there are two elements that must be adressed. Firstly, its likely membership to Sgr~OB1 suggest its mass can be above $10\:\mathrm{M}_{\sun}$ (if directly associated to NGC~6531), which is at odds with the upper mass limit for AGB stars obtained by some authors \citep[e.g.][]{doh2015}. However, models for single stars as massive as $12\:\mathrm{M}_{\sun}$ entering the AGB exist, given choices on semiconvection and convective stability \citep{poelarends08, siess06}. Secondly, the high luminosity we derived for VX~Sgr ($M_{\rm bol}$~$=$~$-8.6\pm0.6$~mag) is above the highest luminosities predicted for AGB stars of 9 to $10\:\mathrm{M}_{\sun}$ \citep{kar2012,doh2015}. Thus, assuming that the mass VX~Sgr is higher than $10\:\mathrm{M}_{\sun}$ would explain its extraordinary high luminosity. The alternative would imply that theoretical models are underestimating the luminosities that AGB stars with slightly below $10\:\mathrm{M}_{\sun}$ can reach. 

\subsection{What drives the variations of VX~Sgr?}
\label{variations}

The mechanisms that drive $T_{\mathrm{eff}}$ variations in VX~Sgr have not been analysed in previous studies. The photometric periodic behaviour has been described as Mira-like pulsations \citep[e.g.][]{hum1974a,gor2018}. This is consistent with our conclusion that VX~Sgr is an AGB star and is also supported by the duplication of atomic lines observed during the second light maximum. We thus confirm the Mira-pulsation nature of the main period of VX~Sgr. Interestingly, line duplication is not seen in any of the two spectra obtained during the first maximum that we observed even in the specra taken with UVES at much higher spectral resolution. In Sect.~\ref{period} we argued that this weak maximum and its precedent minimum belong to the quiescent phase. Therefore, the absence of line duplication during this maximum indicates that the variation during quiescent phases is not driven by shock-wave pulsations, as during active phases.\\

Additional information on the photometric variability of VX~Sgr can be found in the the Diffuse Infrared Background Experiment \citep[DIRBE; ][]{pri2010} data. They observed VX~Sgr on a weekly basis from September~1989 to July~1993 (which corresponds to active phase A3) in the infrared bands at $1.25\:\mu$m, $2.2\:\mu$m, $3.5\:\mu$m, and $4.9\:\mu$m. From these observations, they derived the variation period for each band. They obtained 622\:d for $1.25\:\mu$m and $3.5\:\mu$m bands, whereas the  $3.5\:\mu$m band had a period of 625\:d. In contrast, the reddest band at $4.9\:\mu$m exhibited a significantly longer period of 808\:d, which is similar to the 770\:d period found for the optical light during active phase A3. As circumstellar dust emission becomes significant at wavelengths longer than $\sim3\mu$m \citep{wol1969}, we can expect that the $4.9\:\mu$m band is dominated by the dust emission. Thus, its period is connected to the pulsations, as they affect the circumstellar envelope (by shock-waves and by radiation due to the significant $T_{\rm eff}$ variations). The other three bands, which are likely dominated by the stellar flux, seem not connected to the pulsation but to the same signal which dominates the quiescent phases. This suggests that the process behind the variations seen in the quiescent phases are present during the active phases. However, when the pulsations responsible for the periodic behaviour of the active phases are present, they dominate completely the visual variation,  obscuring the signals from the quiescent-type variability in that band. \\

In a large and evolved star, such as VX~Sgr, the usual suspect for variability when pulsations are absent is the presence of large convective cells \citep{gra2008,sto2010,kra2019}. In fact, the variations of VX~Sgr during quiescent phases, with amplitudes of $\sim$2\:mag in the visual band (see Fig.~\ref{lightcurve}), are remarkably similar to those observed in typical RSGs \citep{kis2006}. This is not surprising as VX~Sgr has a size (between 1000 and $2000\:$R$_{\sun}$, depending on the distance assumed; e.g. \citealt[][]{mon2004,xu2018}) similar to the size of luminous RSGs \citep{arr2015}. Thus, the behaviour of the quiescent phases may well be caused by large convective cells. In fact, the RV curve before the first light-maximum that we covered seems to anticipate it, although this maximum is likely not caused by a pulsation. This relation between the RV and the light curve has also been observed in RSGs and it is believed to be caused by large convective cells \citep[see ][]{kra2019}. However, the data available do not allow a final answer about the nature of the quiescent phases. Further observations during a quiescent phase will be necessary for that.

\subsection{Thermal pulses in VX~Sgr}
\label{thermal}

As we have concluded that VX~Sgr is a very massive TP~AGB star, a possible explanation can be advanced for the signal likely found at $28\,279\:$d. When a thermal pulse takes place the luminosity oscillates significantly \citep[e.g.][]{vas1993,doh2014a,pig2016}. There are variations in luminosity as large as 75\% (1.5~bolometric magnitudes), but also as small as 5\%. As can be seen in Fig.~\ref{lightcurve}, the $28\,279\:$d signal has an amplitude of $\sim1\:$mag, which is compatible with the observable effect of a thermal pulse. The time between individual thermal pulses in AGB stars is highly dependent on the mass of the star \citep{woo1981,kar2012,doh2015}. It is about $10^4\:$a at 5\:M$_{\odot}$, but at the highest masses explored by \citet{doh2015}, above 9.5\:M$_{\odot}$, it can be as short as several decades, which is compatible with the signal of $28\,279\:$d. This also would be in agreement with our previous conclusions about the mass of VX~Sgr being close to the RSG limit.  Interestingly, the smoothed light curve in Fig~\ref{lightcurve} has a near-sinusoidal shape. This shape is different from what is predicted in the literature \citep[see e.g.,][]{woo1981,iben1983} for low-mass AGBs. However, the shapes shown in \citep{woo1981} are heavily dependent on the stellar mass and, in consequence, they should be interpreted with care. In any event, we must remark that this scenario is purely speculative, and the nature of this signal at $28\,279\:$d cannot be proved by our current data.\\

\subsection{Is VX~Sgr a super-AGB star?}
\label{superagb}

We must consider the possibility of VX~Sgr being a super-AGB star. Such objects are  AGB stars that can ignite their carbon cores prior to their TP phase. This ignition depends on several physical parameters, but mainly on the stellar mass and metallicity. According to theoretical models of \citet{doh2015}, at metallicity $Z=0.02$ we should expect a minimum mass of $8\:$M$_{\odot}$ to ignite the carbon core. Since super-AGB stars are expected to possess higher masses, aside from the presence of \ion{Li}{i} or \ion{Rb}{i}, the main observational factor used to propose a super-AGB candidate is the stellar luminosity \citep{loo2017}. However, an enhanced Rb and a high luminosity are not enough to distinguish a high mass AGB star from a super-AGB star \citep{doh2015}. Consequently, no super-AGB has been  fully confirmed \citep{loo2017} to the present date.  On the other hand, \citet{ograd20} have recently reported on a population of super-AGB candidates in the Magellanic Clouds. These authors find these stars to have overall characteristics (i.e. luminosity, mass, pulsation period, ...) compatible with the expected properties of super-AGB stars, provided that they are burning carbon in their cores and have not reached the super-wind phase yet. VX~Sgr shares some characteristics with these objects, but it is brighter than the upper luminosity range of the candidate super-AGB stars reported by \citet{ograd20}. The higher luminosity of VX~Sgr, as well as other differences respect to the candidates found by \citet{ograd20} (presence of Rb, absence of Li, a significantly higher mass-loss), can be easily explained by the higher mass of VX~Sgr, and perhaps also by a more advanced evolutionary stage. Of particular interest is the very luminous star HV~838, in the SMC \citep[SMC-3 in][]{ograd20}. This object presents a huge photometric variability (almost 3~mag in the $I$ band) with a period of $\sim660\:$d and extreme spectral type variability, ranging from late K to M8 \citep{gon2015}. These properties are similar to those of VX~Sgr, with the earlier spectral types displayed attributable to the lower metallicity of the SMC. \citet{ograd20} conclude that HV~838 is most likely a super-AGB star of 8\,--\,$9\:\mathrm{M}_{\sun}$. However, with an estimated luminosity $\log (L_{*}/L_{\sun})\approx4.8$, HV~838 is still significantly fainter than VX~Sgr. In the case of VX~Sgr, the requisites of enhanced Rb and high luminosity are met, and the available hints support the idea that its mass may be above $10\mathrm{M}_{\odot}$. In this situation, with such a high mass, the probability of VX~Sgr being a super-AGB is as high as it can be from a theoretical standpoint. Thus, we think that VX~Sgr can be considered  the first strong candidate to super-AGB star in the Milky Way.  \\

\section{Conclusions}
\label{conclusions}

In this work, we have calculated the stellar atmospheric parameters and radial velocities for VX~Sgr on 11 different epochs. We summarise the conclusions below:  

\begin{enumerate}

\item The presence of Rb together with the line-doubling during the maximum light confirm that VX~Sgr is not a RSG. As it lacks strong line absorption due to \ion{Li}{i} or \ion{Ca}{i}, features expected in a T\.ZO, we conclude that it is an AGB star. If so, its observed characteristics imply that its initial mass is high, at least above~$7\:{\rm M}_{\odot}$. Moreover, the extraordinary luminosity of VX~Sgr combined with its likely membership to Sgr~OB1 suggest that its mass is extraordinarily high (between $10$ and $12\:\mathrm{M}_{\odot}$). This makes VX~Sgr a strong candidate to super-AGB star.\\

\item Most of the distances calculated for VX~Sgr result in $M_{\mathrm{bol}}$ brighter than the most luminous AGB stars already identified \citep[$M_{\rm bol}\sim$~$-8$~mag;][]{gro2009, gar2009}. In fact, the luminosity for the average of distance estimates, $M_{\mathrm {bol}}\sim-8.6\:$~$\pm$~0.6 mag, is clearly above that limit. Thus, we conclude that VX~Sgr actually is the most luminous AGB known to date.\\

\item The line-doubling of atomic lines during the second maximum that we observed, confirms the Mira-like pulsational nature of the variations seen in VX~Sgr during active phases. The presence of emission in the Balmer lines at maximum light is also typical of this type of stars. On the other hand, the absence of this duplication during the previous maximum, the last peak in the previous quiescent phase, suggests that the pulsation of VX~Sgr is not the drive behind the variations seen during quiescent phases. This second driving mechanism is working during both active and quiescent active phases, but is obscured by the much more powerful pulsation during the former. After analysing the data available, we think that convective cells are a solid candidate to be this drive.\\

\item We have computed stellar parameters and SpTs for VX~Sgr and found a good agreement between both of them. More observations are required to correctly sample these late SpTs, although we consider that a good correlation exists, given that we only have 11 points to assess  it. More points will be required to properly sample the $T_{\rm eff}$ scale at these late SpTs, as demonstrated by \citet{tab18}.\\

\item We observed two epochs in which we see a photospheric \ion{Rb}{i} component, whereas a blueshifted circumstellar component is always present in both \ion{Rb}{i} lines. This is the first time that such behaviour is observed in an evolved star. The variation of these lines implies strong changes in the circumstellar envelope, probably related to the end of the last and extraordinarily long quiescent phase. The circumstellar components display expansion velocities compatible with those observed in OH masers. Such outflow requires a regular supply of s-process elements to the stellar atmosphere.\\

\item  We calculate a main period of 757~d, similar to the value obtained by \cite{kam2005}. However, this main period is product of the overlapping between the different main periods of the different phases that VX~Sgr has been going through. In addition, we found a signal at $28\,279\:$d ($\sim77\:$a)  which provides a good overall fit to the AAVSO light curve. To our best knowledge, this signal has never been reported before. Although the data available are insufficient to demonstrate whether this is a true periodic behaviour, we speculate that this variation may be connected with the thermal pulses expected for a very massive AGB star.\\

\end{enumerate}

\begin{acknowledgements}
We thank Dr.~D.~An\'ibal Garc\'ia-Hern\'andez for our fruitful conversations about VX~Sgr and the advice about the methodology to analyse AGB spectral data.
We gratefully acknowledge the variable star observations from the AAVSO International Database with contributions of observers worldwide and used in this research. This work has made use of data from the European Space Agency (ESA) mission {\it Gaia} (\url{https://www.cosmos.esa.int/gaia}), processed by the {\it Gaia}  Data Processing and Analysis Consortium (DPAC, \url{https://www.cosmos.esa.int/web/gaia/dpac/consortium}). Funding for the DPAC has been provided by national institutions, in particular the institutions participating in the {\it Gaia} Multilateral Agreement. This research is partially supported by the Spanish Government Ministerio de Ciencia e Innovaci\'on (MICI) under grants FJCI-2014-23001, AYA2015-68012-C2-2-P, PGC2018-093741-B-C21/C22 (MICI/AEI/FEDER, UE). This work was also supported by Fundação para a Ciência e a Tecnologia (FCT) through the research grants UID/FIS/04434/2019, UIDB/04434/2020 and UIDP/04434/2020. HMT also acknowledges support from the FCT - Fundação para a Ciência e a Tecnologia through national funds (PTDC/FIS-AST/28953/2017) and by FEDER - Fundo Europeu de Desenvolvimento Regional through COMPETE2020 - Programa Operacional Competitividade e Internacionalização (POCI-01-0145-FEDER-028953). RD acknowledges support from the Spanish Government Ministerio de Ciencia e Innovaci\'on (MICI) through grants PGC-2018-091\,3741-B-C22 and SEV 2015-0548, and from the Canarian Agency for Research, Innovation and Information Society (ACIISI), of the Canary Islands Government, and the European Regional Development Fund (ERDF), under grant with reference ProID2017010115. EM acknowledges financial support from the Spanish Ministerio de Ciencia e Innovación through fellowship FPU15/01476.
\end{acknowledgements}
\bibliographystyle{aa} 
\bibliography{VXSgr}
\begin{appendix}
\section{Extra material}
 \begin{figure*}                                                  \centering                       
    \includegraphics[width=\textwidth]{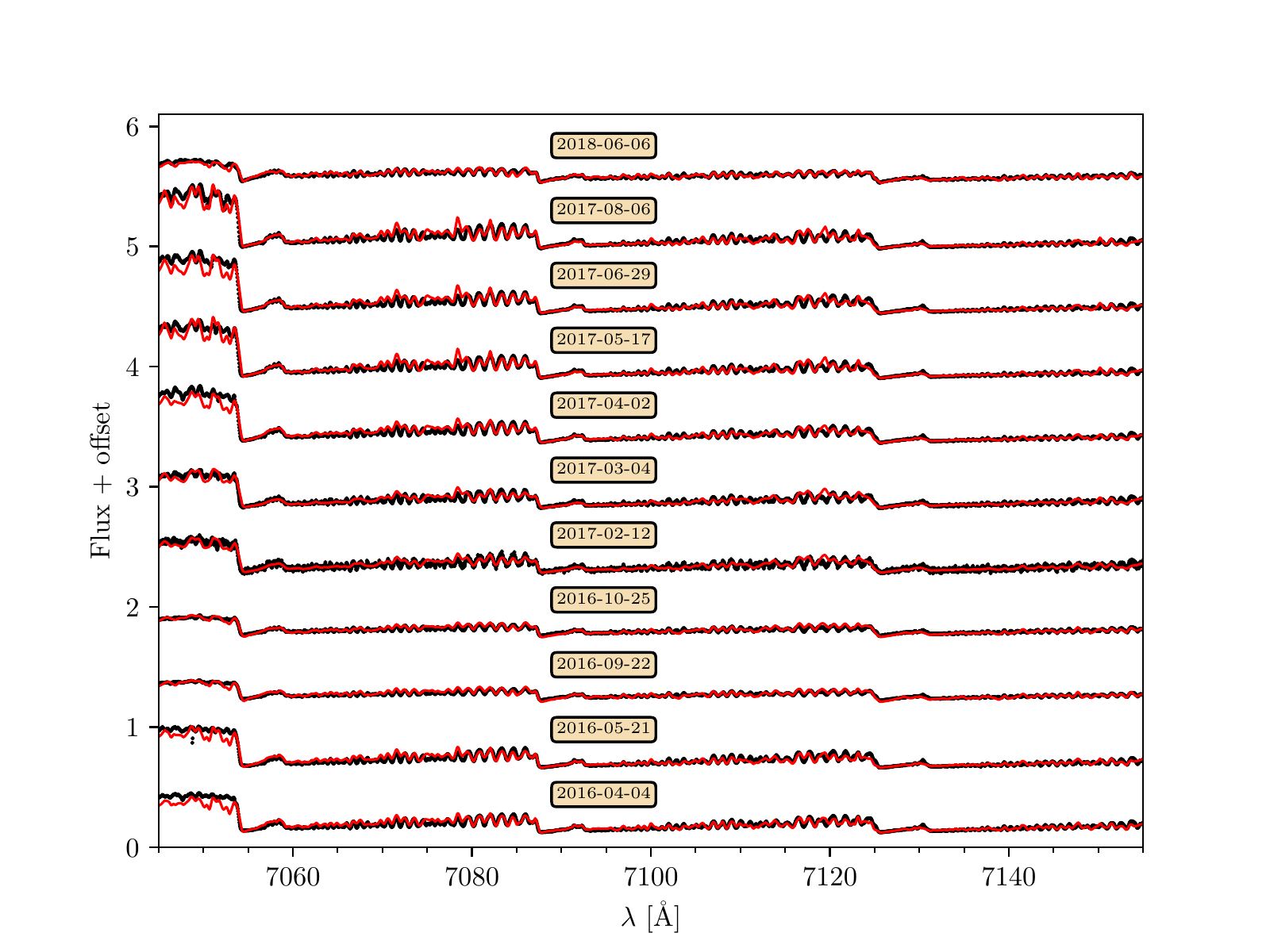}
    \caption{Best fit PHOENIX-ACES models (red) to our spectroscopic observations (black).}
    \label{best_fit}
\end{figure*}

 \begin{figure*}                                                  \centering                       
    \includegraphics[scale = 0.5]{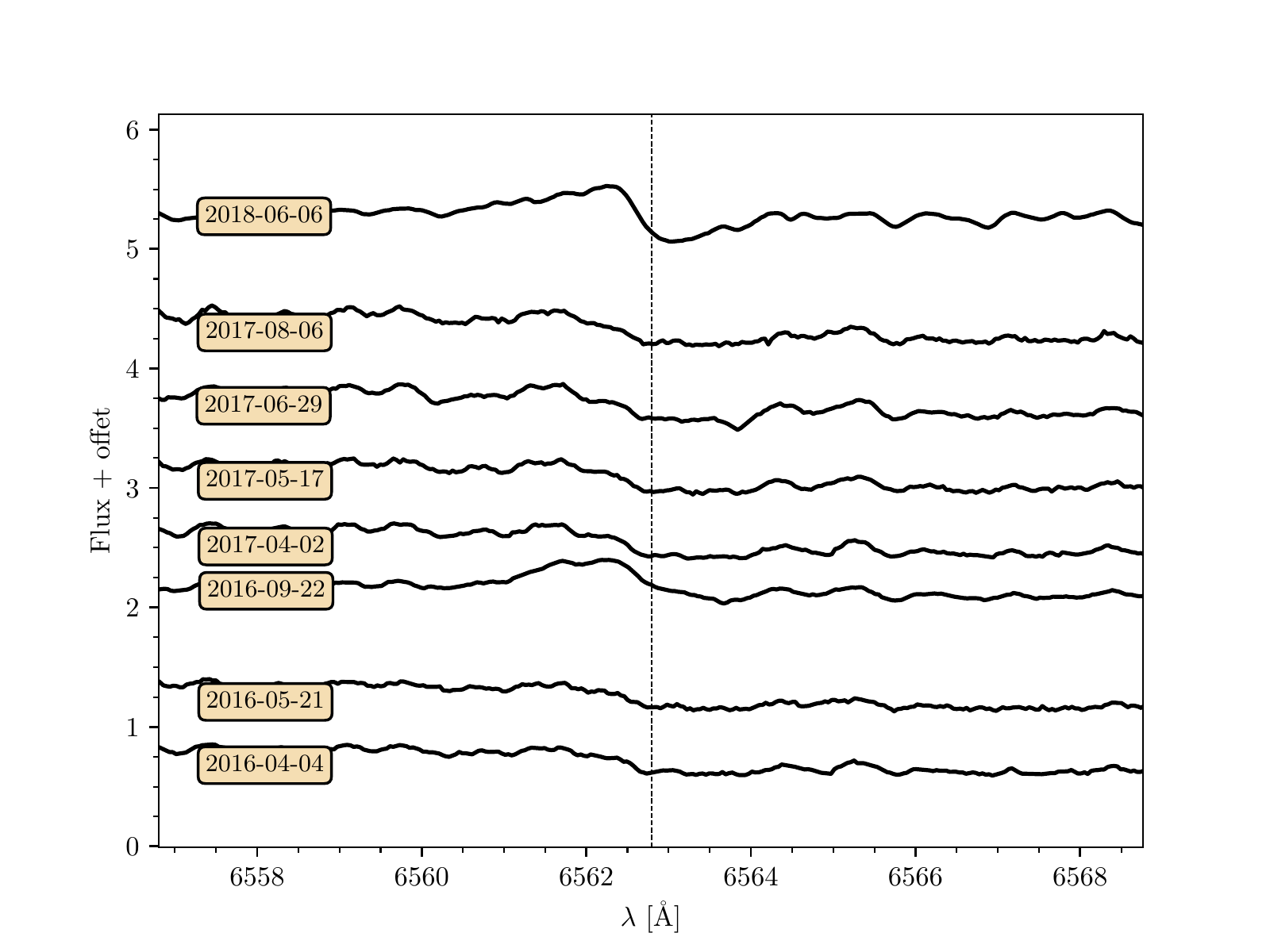}
    \includegraphics[scale = 0.5]{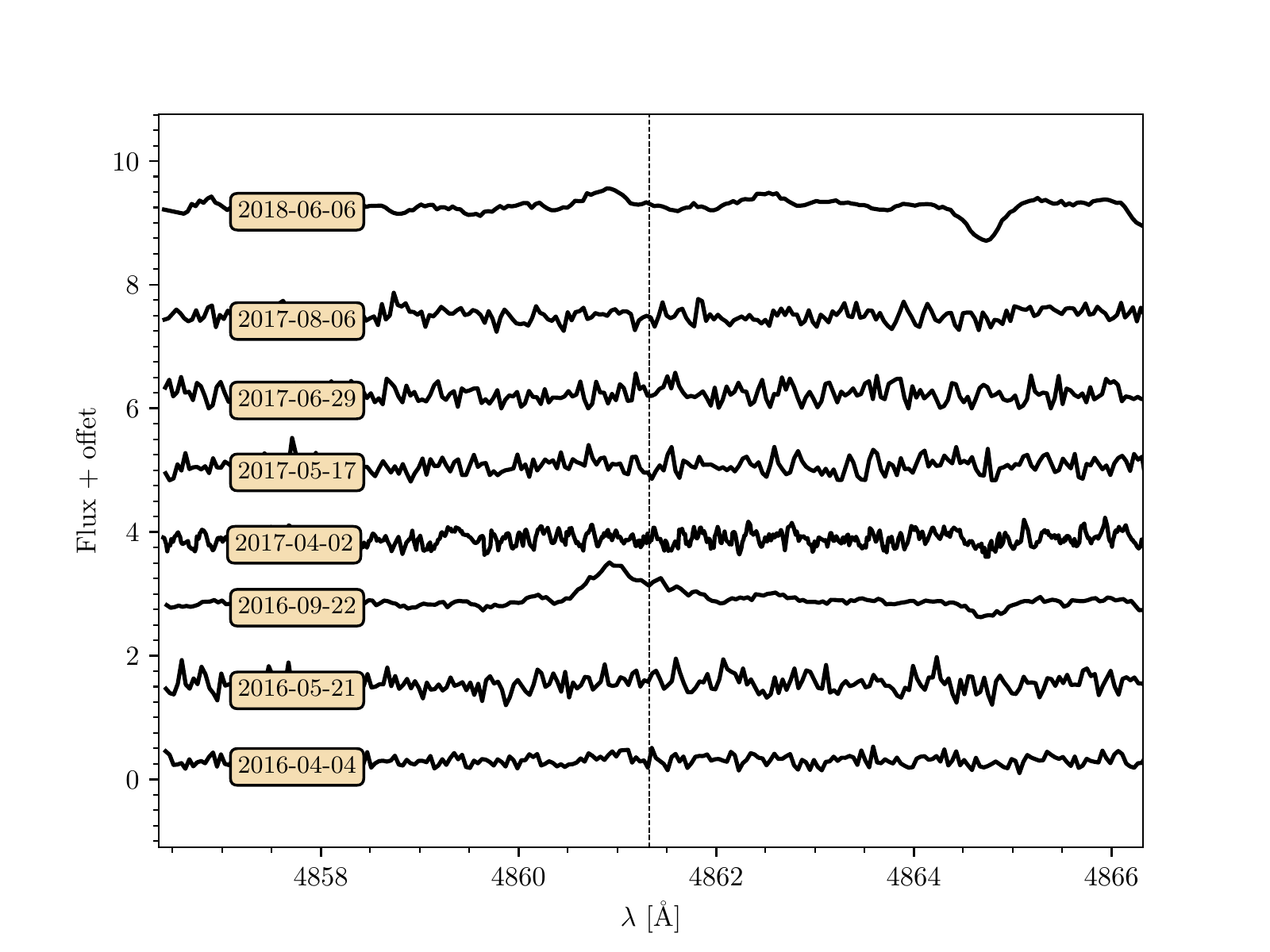}    
    \includegraphics[scale = 0.5]{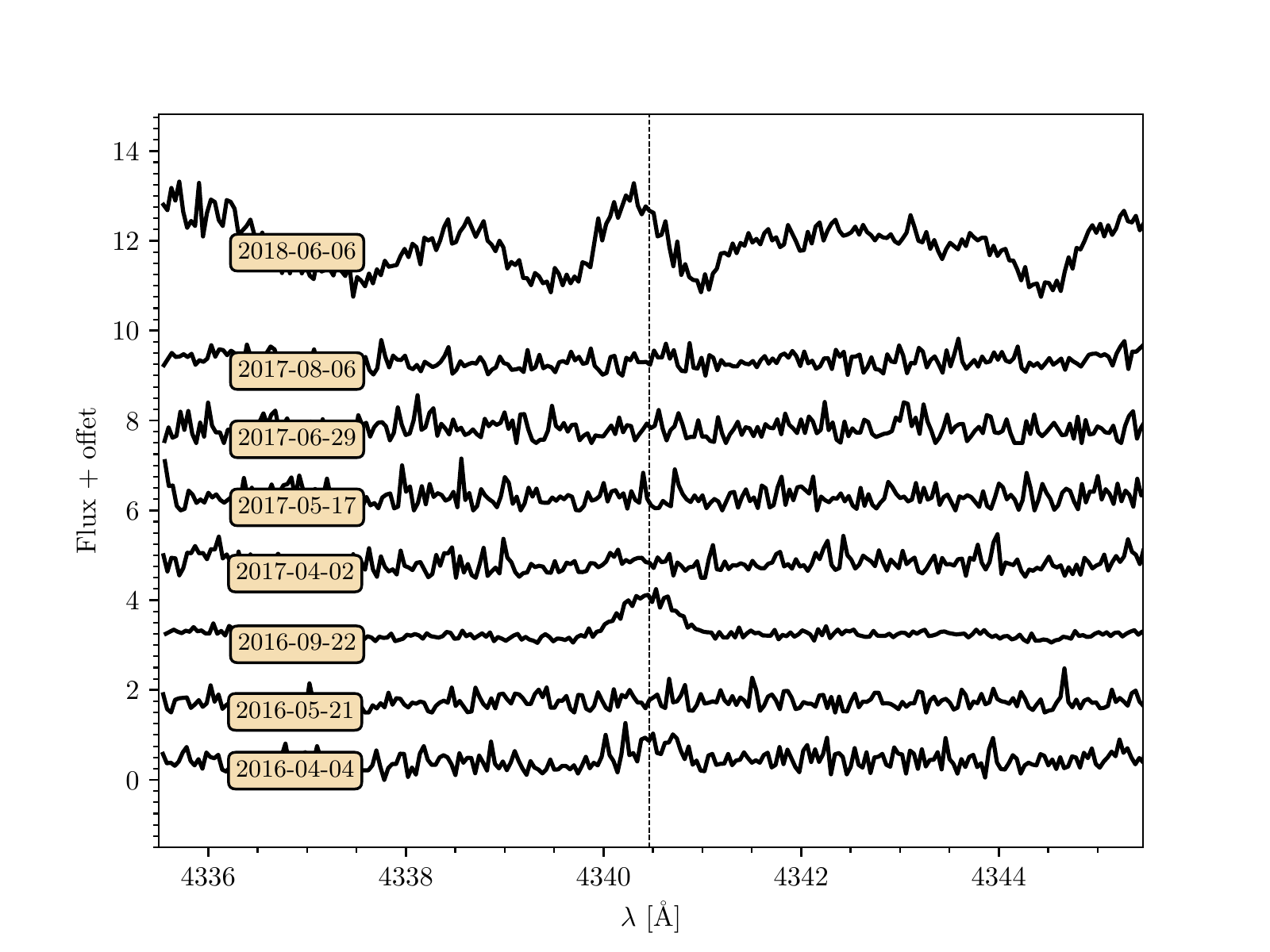}    
    \includegraphics[scale = 0.5]{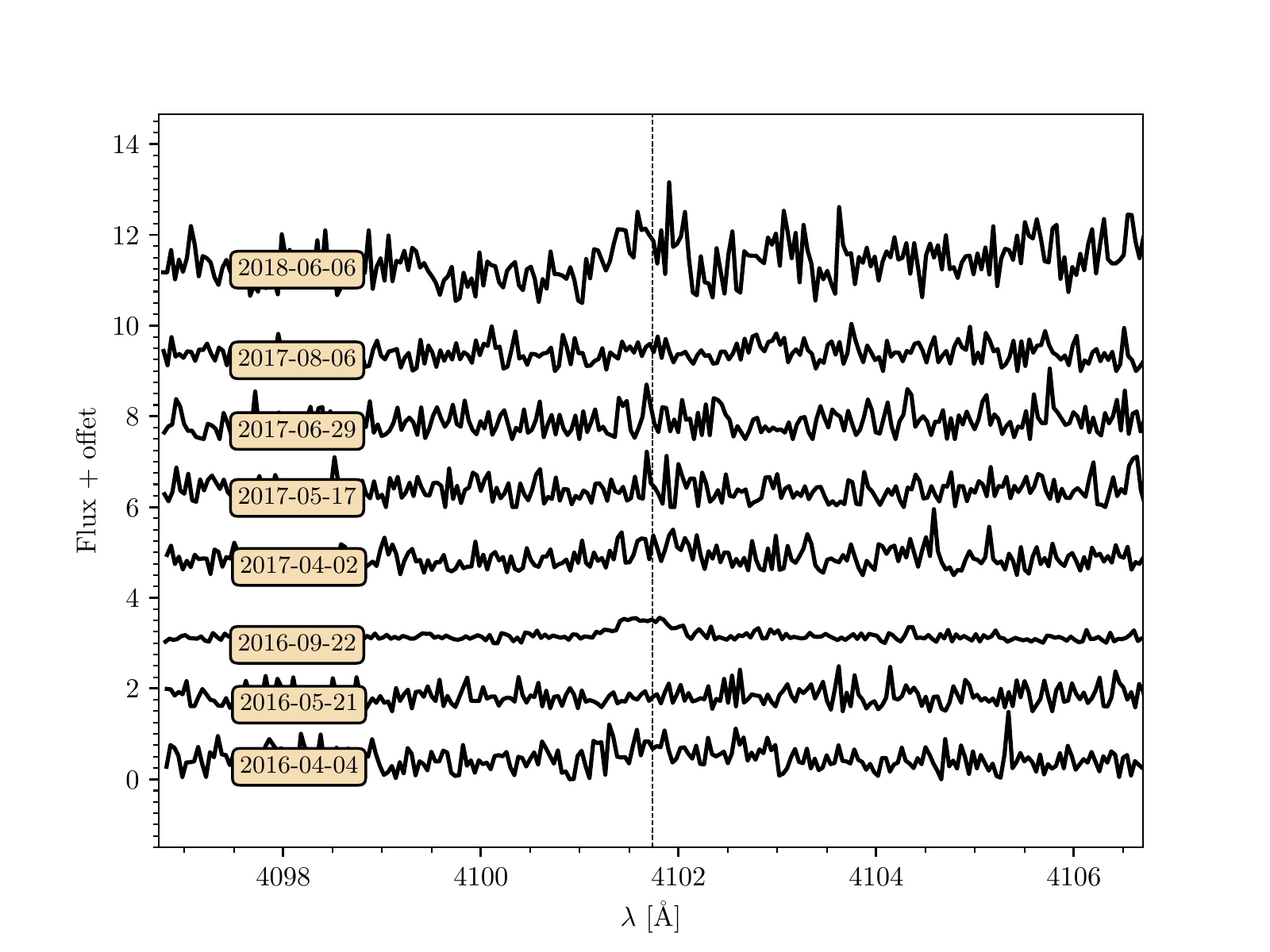}    
    \caption{VX~Sgr around the lines of the Balmer series (from H$\alpha$ to H$\delta$).}
    \label{Balmer}
\end{figure*}

\end{appendix}
\end{document}